\documentclass[twocolumn]{aastex7}
\received{December 4, 2025}
\revised{January 28, 2026}
\accepted{February 13, 2026}
\usepackage{placeins} 
\usepackage{amssymb}
\usepackage{amsmath}
\usepackage{pdfrender}
\usepackage{caption}
\newcommand*{\boldcheckmark}{%
  \textpdfrender{
    TextRenderingMode=FillStroke,
    LineWidth=.5pt, 
  }{\checkmark}%
}
\begin{document}

\title{Probing AGN Feedback in Dwarf Galaxies with Spatially Resolved NIR Coronal Lines from JWST}

\author[0000-0001-7578-2412]{Archana Aravindan}
\affiliation{Department of Physics and Astronomy, University of California, Riverside, 900 University Ave, Riverside CA 92521, USA}
\affiliation{Department of Astronomy, The University of Texas at Austin, Austin, TX 78712, USA}
\affiliation{Cosmic Frontier Center, The University of Texas at Austin, Austin, TX 78712, USA}
\email[show]{archana.aravindan@austin.utexas.edu}

\author[0000-0002-4375-254X]{Thomas Bohn}
\affiliation{Ehime University, Bunkyo-cho 2-5, Matsuyama, Ehime 790-8577, Japan}
\email{thomas.bohn@email.ucr.edu}

\author[0000-0003-4693-6157]{Gabriela Canalizo}
\affiliation{Department of Physics and Astronomy, University of California, Riverside, 900 University Ave, Riverside CA 92521, USA}
\email{gabyc@ucr.edu}

\author[0000-0003-2277-2354]{Shobita Satyapal}
\affiliation{Department of Physics and Astronomy, George Mason University, MS3F3, 4400 University Drive, Fairfax, VA 22030, USA}
\email{}

\author[0000-0002-1912-0024]{Vivian U}
\affiliation{IPAC, California Institute of Technology, 1200 E. California Blvd., Pasadena, CA 91125, USA}
\email{}

\author[0000-0003-3762-7344]{Weizhe Liu}
\affiliation{Steward Observatory, University of Arizona, 933 N. Cherry Ave,
Tucson, AZ 85721, USA}
\email{}

\author[0000-0003-3937-562X]{William Matzko}
\affiliation{Department of Physics and Astronomy, George Mason University, MS3F3, 4400 University Drive, Fairfax, VA 22030, USA}
\email{}

\author[0000-0003-3152-4328]{Sara Doan}
\affiliation{Department of Physics and Astronomy, George Mason University, MS3F3, 4400 University Drive, Fairfax, VA 22030, USA}
\email{}

\author[0000-0001-6919-1237]{Matthew Malkan}
\affiliation{Division of Astronomy and Astrophysics, University of California, Los Angeles, CA 90095}
\email{}

\author[0000-0003-3498-2973]{Lee Armus}
\affiliation{IPAC, MC 320-6, Caltech, 1200 E. California Blvd., Pasadena, CA 91125}
\email{}

\author[0000-0002-7402-5441]{Tohru Nagao}
\affiliation{Ehime University, Bunkyo-cho 2-5, Matsuyama, Ehime 790-8577, Japan}
\email{}

\author[0000-0003-0699-6083]{Tanio Diaz-Santos}
\affiliation{Institute of Astrophysics - FORTH, GR-70013 Vassilika Vouton,
Greece}
\email{}

\author[0009-0003-7749-1864]{Aditya Togi}
\affiliation{Department of Physics, Texas State University, 601 University Drive, San Marcos, TX 78666, USA}
\email{}

\author[0000-0001-8490-6632]{Thomas S.Y. Lai}
\affiliation{IPAC, MC 320-6, Caltech, 1200 E. California Blvd., Pasadena, CA 91125}
\email{}

\author[0000-0002-1000-6081]{Sean T. Linden}
\affiliation{Steward Observatory, University of Arizona, 933 N. Cherry Ave,
Tucson, AZ 85721, USA}
\email{}

\author[0000-0002-6570-9446]{Marina Bianchin}
\affiliation{Instituto de Astrof\'{\i}sica de Canarias, Calle V\'{\i}a L\'{a}ctea s/n, E-38205, La Laguna, Tenerife, Spain}
\affiliation{Departamento de Astrof\' isica, Universidad de La Laguna, E-38206, La Laguna, Tenerife, Spain}
\email{}

\author[0000-0002-3139-3041]{Yiqing Song}
\affiliation{European Southern Observatory, Joint ALMA Observatory, Alonso de Co\'{r}dova, 3107, Vitacura, Santiago, 763-0355, Chile}
\email{}

\author[0000-0003-0057-8892]{Loreto Barcos-Muñoz}
\affiliation{North American ALMA Science Center, National Radio Astronomy Observatory, 520 Edgemont Road, Charlottesville, VA 22903}
\email{}

\author[0000-0003-2638-1334]{Aaron Evans}
\affiliation{Department of Astronomy, University of Virginia, 530 McCormick Road, Charlottesville, VA 22904
NRAO, 520 Edgemont Road, Charlottesville, VA, 22903}
\email{}

\author[0000-0003-4268-0393]{Hanae Inami}
\affiliation{Hiroshima University, 1-3-1 Kagamiyama, Higashi-Hiroshima City, Hiroshima, 739-8526, Japan}
\email{}

\author[0000-0003-3917-6460]{Kirsten Larson}
\affiliation{Space Telescope Science Institute, Baltimore, MD 21218}
\email{}

\author[0000-0002-7532-3328]{Sabrina Stierwalt}
\affiliation{Occidental College, 1600 Campus Road, Los Angeles, CA, 90041}
\email{}

\author[0000-0001-7291-0087]{Jason Surace}
\affiliation{IPAC, MC 314-6, Caltech, 1200 E. California Blvd., Pasadena, CA 91125}
\email{}

\begin{abstract}
We present the first spatially resolved investigation of near-infrared coronal lines in dwarf galaxies hosting active galactic nuclei (AGN), using JWST/NIRSpec integral field spectroscopy. Coronal lines (CLs), which are forbidden transitions from highly ionized species with ionization potentials up to 450 eV, act as sensitive tracers of the AGN ionizing continuum and feedback processes. Across four dwarf galaxies with ionized gas outflows traced by the optical [O\,III] lines, we report the detection of 16 unique species of near-infrared CLs. Line ratio diagnostics indicate that photoionization from the AGN dominates the excitation of CLs.  We find that the coronal line region in dwarf galaxies, traced by the various CLs, extends up to 0.5 kpc, and can constitute up to 10\% of their host galaxy size. Correlations between CL luminosities and [O\,III] ionized gas outflow properties are consistent with a scenario in which AGN-driven outflows likely facilitate the detection of CLs and contribute to their extent. Several CLs, including [Si\,VI], [Si\,VII], and [Mg\,VIII], exhibit a secondary broad component with W$_{80}$ (the line width enclosing 80\% of the total flux) $>$ 300 km s$^{-1}$.  If we interpret this spatially compact gas as part of an outflow, this would indicate that the outflowing gas includes a wide range of ionizations. The estimated energetics imply this highly ionized component is compact yet powerful enough to perturb gas in the central regions of the host dwarfs. These results indicate that AGN in low-mass galaxies may produce outflows capable of influencing their structure and evolution.

\end{abstract}


\section{Introduction} 

Over the past decade, a growing number of studies have revealed that active galactic nuclei (AGN) are not limited to massive galaxies, but also occur in dwarfs with stellar masses below $\sim 10^{9} M_{\odot}$ \citep{2003ApJ...588L..13F, 2013ApJ...775..116R, 2014AJ....148..136M, 2018ApJ...863....1C, 2020ApJ...898L..30M, 2022ApJ...937....7S, 2025ApJ...982...10P}. Their presence in such low-mass systems provides new constraints on black hole seed formation and the role of feedback in shallow gravitational potentials \citep{2020ARA&A..58...27I, 2021NatRP...3..732V, 2021MNRAS.503.3568K, 2022MNRAS.516.2112K, 2023ApJ...946...51C}. Traditionally, stellar feedback from supernovae and young stars has been regarded as the dominant regulator of star formation and gas loss in dwarfs, but recent observations show that AGN can also drive galaxy-wide outflows \citep{2018MNRAS.476..979P, 2019ApJ...884...54M, 2024ApJ...965..152L, 2025ApJ...979...26S, 2025A&A...697A.235R}. In some cases, the mass and kinetic energy outflow rates of AGN in dwarf galaxies are comparable to those in massive galaxies when scaled by AGN luminosity \citep{2019ApJ...884...54M, 2020ApJ...905..166L}, and AGN-driven winds have been shown to be faster and more energetic than stellar-driven outflows in hosts of similar mass and at similar redshifts \citep{2023ApJ...950...33A}.

\begin{deluxetable*}{ccccccccc}
\tablecaption{Target Details}
\tablehead{
\colhead{Name} & \colhead{Short name} & \colhead{Redshift} & \colhead{log(M$_*$/M$_{\odot}$)} & \colhead{log(M$_{BH}$/M$_{\odot}$)} & \colhead{log[L(AGN)]}  & \colhead{SFR} & \colhead{v$_{50, \, \mathrm{[O\,III]\,broad}}$} & \colhead{W$_{80, \, \mathrm{[O\,III]\,broad}}$}\\
\colhead{} & \colhead{} & \colhead{} & \colhead{} & \colhead{} & \colhead{(erg s$^{-1}$)}  & \colhead{(M$_{\odot}$ yr$^{-1}$)} & \colhead{(km s$^{-1}$)} & \colhead{(km s$^{-1}$)}\\
\colhead{(1)} & \colhead{(2)} & \colhead{(3)} & \colhead{(4)} & \colhead{(5)} & \colhead{(6)} & \colhead{(7)} & \colhead{(8)} & \colhead{(9)}}
\startdata
J084234.51+031930.7 & J0842 & 0.0291 & 9.3 & 5.8 $\pm$ 0.5 (i) & 43.1 $\pm$ 0.4  &  $<$0.3 & -110 & 500\\
J090613.75+561015.5 & J0906 & 0.0467 & 9.4 & 5.4 $\pm$ 0.3 (ii) & 43.7 $\pm$ 0.8  & $<$0.3 & -70 & 980\\
J095418.16+471725.1 & J0954 & 0.0329 & 9.1 & 5.0 $\pm$ 0.4 (ii) & 43.9 $\pm$ 1.1  & $<$0.3 & -60 & 730\\
J100935.66+265648.9 & J1009 & 0.0145 & 8.8 & 5.1 $\pm$ 0.1 (iii) & 43.0 $\pm$ 1.0  & $<$0.1 & -20 & 210\\
\enddata
\tablecomments{\footnotesize
Columns (1): SDSS name of the target. (2): Short name used in this paper. (3): Redshift, as measured from a combination of stellar absorption features and Pa$\alpha$ 1.86 $\mu$m. (4): Stellar mass from the NASA Sloan Atlas \citep{2011AJ....142...31B} with an average error of 0.5 dex. (5): Black hole mass derived using broad line fits and Eddington luminosities from the following references: (i) \citealt{2019ApJ...884...54M}, (ii) \citealt{2013ApJ...775..116R, 2017AA...602A..28M}, (iii) \citealt{2014AJ....148..136M}. We caution that the estimated masses are uncertain, by 0.5 dex or more. (6): Bolometric AGN luminosity based on extinction-corrected [O\,III] luminosity \citep{2020ApJ...905..166L, 2024ApJ...965..152L}. (7): Upper limit on the star formation rate (SFR) based on the extinction-corrected [O\,II] $\lambda \lambda$3726, 3729 from \citealt{2020ApJ...905..166L}, assuming that one-third of the [O\,II] emission is from the star formation activity, following
\citealt{2005ApJ...629..680H}. (8): Median v$_{50}$ (velocity at the 50th percentile of the total velocity) measurements of the broad, outflowing component in [O\,III] determined from KCWI data \citep{2020ApJ...905..166L}. (9): Median W$_{80}$ (line width that includes 80\% of the total flux)) measurements of the broad, outflowing component in [O\,III] also determined from KCWI data \citep{2020ApJ...905..166L}.}
\end{deluxetable*}\label{tab:tab1}

Unlike stellar feedback, which mostly acts locally and is limited by the star formation rate, AGN can inject energy on galaxy-wide scales, often with outflow velocities that exceed the escape velocity of dwarfs \citep{2019ApJ...884...54M}. Even short-lived AGN episodes may therefore have disproportionate effects compared to continuous stellar activity, regulating star formation, removing gas and metals, and enriching the circumgalactic medium. If AGN feedback proves effective at these mass scales, it would imply that black hole growth and galaxy evolution are coupled across the full mass spectrum, challenging the paradigm that stellar feedback dominates in dwarfs while AGN feedback shapes massive galaxies \citep{2011IAUS..277..273S}. This raises a key question: can AGN feedback play an equally significant role in dwarf galaxy evolution as in their massive counterparts?

A major obstacle in answering this question lies in confirming that the observed outflows are indeed AGN-driven. Common tracers of ionized gas such as [O\,III] $\lambda$5007 can suffer contamination from stellar processes, making it challenging to distinguish between stellar- and AGN-powered winds in these low-mass systems. This necessitates the need for observational diagnostics that are both sensitive to AGN photoionization and largely immune to stellar contamination \citep{2019ApJ...870L...2C}.

Coronal lines (CLs), which are forbidden transitions from highly ionized species with ionization potentials up to 450 eV, satisfy these requirements. Their excitation typically requires the hard radiation field from an accreting black hole. Although recent observations have detected CLs in stellar winds from massive stars \citep{2024ApJ...976L..25L}, the presence of highly ionized ($>$150 eV) and extended CLs are unlikely to be produced by stellar populations alone \citep{2021ApJ...906...35S, 2022ApJ...927..165R}. Thus, CLs provide a unique window into the otherwise unobservable ionizing continuum of Type-2 AGN, and have been used to constrain the spectral energy distribution of the central engine, estimate black hole masses \citep{2018ApJ...861..142C, 2022MNRAS.510.1010P}, and trace the most highly ionized phases of AGN-driven outflows \citep{2011ApJ...739...69M, 2018MNRAS.481L.105M, 2022ApJ...940L...5U, 2023MNRAS.524..143F, 2023ApJ...942L..37A, 2024ApJ...965..103B}. Their high ionization potentials ensure that the emission is free from contamination by stellar outflows. In massive galaxies, CLs have been observed to extend beyond the narrow-line region, often tracing interactions between AGN jets and the interstellar medium \citep{2006ApJ...653.1098R, 2021ApJ...920...62N, 2025ApJ...984..170M}. Because CLs likely trace only the AGN-powered, most highly ionized phases of outflows, they provide a direct probe of AGN feedback. This is particularly critical in dwarf galaxies, where the shallow potential wells make them more susceptible to the gas-clearing effects of even modest AGN-driven winds \citep{2017ApJ...839L..13S}.

Despite their potential to identify AGN and AGN-feedback in more massive galaxies, CLs in dwarf galaxies remain largely unexplored. Photoionization modeling predicts that intermediate-mass black holes (IMBHs) in dwarfs should produce stronger CL emission due to their harder ionizing spectral energy distribution (SED) \citep{2018ApJ...861..142C}. Yet, only a small number of dwarf galaxies have reported detections of CLs, both in the optical and near-infrared \citep{2021ApJ...911...70B, 2021ApJ...912L...2C, 2021ApJ...922..155M, 2023ApJ...946L..38R, 2024ApJ...975...60A}. Ground-based NIR spectroscopy of dwarfs hosting AGN-driven, ionized-gas outflows has revealed high-ionization CLs \citep{2021ApJ...911...70B}, but their spatial extents, kinematics, and connection to AGN feedback remain poorly constrained.

In this paper, we use JWST/NIRSpec integral field spectroscopy to spatially map the coronal line region (CLR) in dwarf galaxies hosting AGN and to examine its relationship to AGN-driven outflows. Our resampled 0.05$\arcsec$ pixel scale of NIRSpec allows us to resolve the inner structure of the CLR, probe the spatial extent of NIR CLs, including several transitions at wavelengths $>$ 2.4 $\mu$m, which are difficult to access from the ground, and confirm the AGN origin of observed outflows. By directly linking CL emission to feedback processes, we provide new constraints on the physical conditions, energetics, and influence of AGN in the low-mass regime. The paper is organized as follows: in Section \ref{sec:Observations and Data reduction}, we present our JWST observations and data reduction; Section \ref{sec:analysis} describes our analysis and spatial mapping techniques; Section \ref{sec:results} presents our CL detections and the ionizing source; Section \ref{sec:discussion} discusses the coronal line region and implications for AGN feedback in dwarf galaxies; and Section \ref{sec:summary} summarizes our findings. Throughout, we assume a flat $\Lambda$CDM cosmology with $H_0$ = 70 km s$^{-1}$ Mpc$^{-1}$, $\Omega_m$= 0.3, and $\Omega_{\Lambda}$ = 0.7.

\section{Observations and Data reduction} 

\subsection{Sample selection: Presence of ionized gas outflows} \label{sec:OIII outflows}

\begin{figure}
     \centering
     \includegraphics[width=1\linewidth]{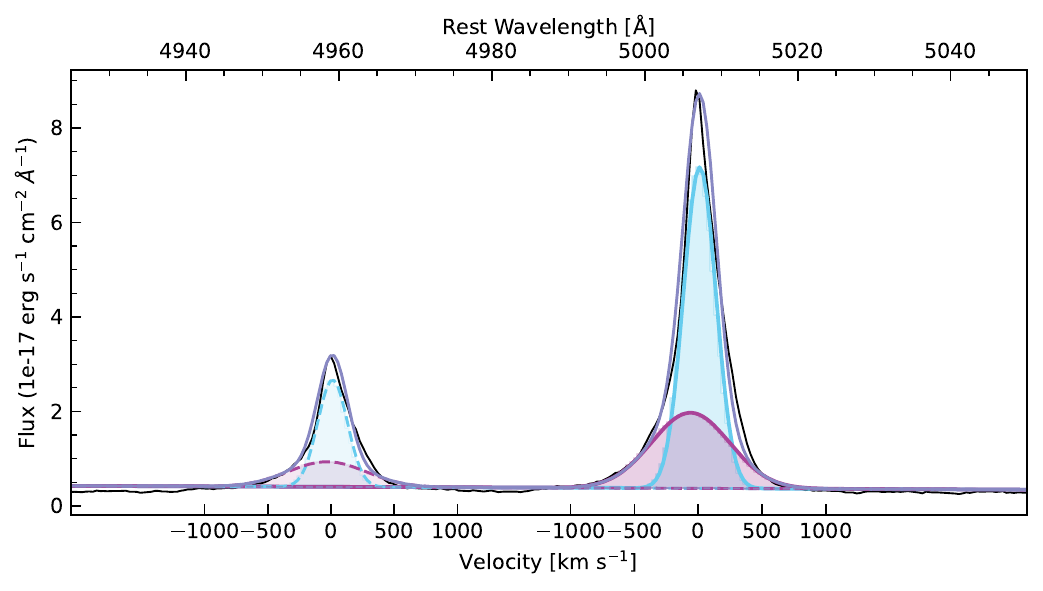}
     \caption{Keck/KCWI spectrum of J0906, one of the targets in the sample, showing the blue-shifted wings of [O\,III] within the central 0.5 kpc. All the targets observed with JWST exhibit similar ionized gas outflow profiles detected in [O\,III].}
     \label{fig:OIII outflows}
\end{figure}

We began with 29 Baldwin–Phillips–Terlevich (BPT; \citealt{1981PASP...93....5B})-selected, spatially extended, AGN-hosting dwarf galaxies \citep{2013ApJ...775..116R,2014AJ....148..136M,2019ApJ...884...54M}. Follow-up Keck/LRIS long-slit spectroscopy and Keck/KCWI IFU data isolated eight systems with galaxy-scale warm-ionized outflows traced by [O\,III] $\lambda\lambda 4959,5007$ doublet \citep{2019ApJ...884...54M,2020ApJ...905..166L} (Figure \ref{fig:OIII outflows}). Near-IR spectroscopy with Keck/NIRES then detected CLs in five of these eight \citep{2021ApJ...911...70B}. From this set, we selected four targets listed in Table \ref{tab:tab1} with stellar masses $<$ 10$^{9.5}$ M$_\odot$ for JWST/NIRSpec IFU that sample a range in stellar mass. Further details of the JWST sample will be provided in T.\ Bohn et al.\ 2025, in prep.

The [O\,III] emission profile in these galaxies has a broad, often blue-shifted wing. \citealt{2019ApJ...884...54M, 2020ApJ...905..166L} model this with an additional Gaussian distinct from the narrow systemic gas with median bulk velocities of  $\sim240~\mathrm{km\,s^{-1}}$ and W$_{80}$ up to $\sim1200~\mathrm{km\,s^{-1}}$. HST/COS observations of three of the fastest systems \citep{2024ApJ...965..152L} show definite outflows, producing blue-shifted UV absorption in the C\,II, C\,IV, Si\,II, and Si\,IV lines. Photoionization modeling for one of the galaxies, J0954, places the absorbing gas at $\sim$ 0.5 kpc, suggesting a galactic-scale impact. The measured kinetic-energy outflow rates also exceed those of starburst winds at comparable SFRs \citep{2024ApJ...965..152L}. Keck/NIRES detections of [Si\,VI] (and other NIR CLs) often show higher outflow velocities than [O\,III], and the lack of J-band CN absorption features argues against a dominant young-stellar origin \citep{2021ApJ...911...70B}. The presence of the AGN makes it challenging to measure direct (T$_e$-based) gas-phase metallicities for our targets. A systematic analysis of AGN-hosting dwarf galaxies in the \citealt{2013ApJ...775..116R} sample, which includes the targets studied in this paper, finds a mean metallicity of $\sim$0.3\,Z$_\odot$ for the sample as a whole \citep{2020ApJ...903...58C}, although individual galaxy values are not reported. Additionally, BPT-selected AGNs in dwarf galaxies are expected to be biased toward higher metallicities \citep{2013ApJ...775..116R}. The relatively high stellar masses of the galaxies in this sample make extremely low metallicities ($\sim$0.01\,Z$_\odot$) unlikely. Throughout this work, we therefore adopt near-solar metallicities, consistent with prior studies of this sample \citep{2020ApJ...905..166L, 2021ApJ...911...70B}, and we also explore trends assuming sub-solar values (0.5\,Z$_\odot$ and 0.25\,Z$_\odot$).

\subsection{JWST/NIRSpec Observations and Data Reduction} \label{sec:Observations and Data reduction}
The data were obtained from the JWST Cycle 2 program (PID 3663, P.I.: Thomas Bohn). All four targets were observed with NIRSpec IFS \citep{ 2022A&A...661A..80J, 2023PASP..135c8001B}. J0906 was observed on 17 March 2024, J1009 on 9 May 2024, J0954 on 8 March 2024 and J0842 on 7 April 2024. Data were collected using the NIRSpec IFU high resolution mode in the G140H/F100LP, G235H/F170LP, and G395H/F290LP grating/filter combinations. The resulting wavelength coverage obtained with our choice of grating/filter combination was 0.9 - 5.2 $\mu$m with a nominal resolving power of R $\sim$ 2700 (110 km s$^{-1}$). We obtained dedicated background observations for the four targets instead of the standard LeakCal exposures in order to subtract the background to detect any faint CL. 

We downloaded the raw data files from the Barbara Mikulski Archive for Space Telescopes (MAST) and subsequently processed them with the JWST Science Calibration pipeline version 1.15.1 \citep{2023zndo...7577320B} under the Calibration Reference Data System (CRDS) context jwst\_1253.pmap. The level 1 files downloaded from MAST were processed using the Detector1 stage (stage 1), which performs detector-level corrections and generates count-rate images. In order to remove artifacts generated by cosmic-ray hits on the detector, we applied the snowball flagging for the jump step during the first stage of the pipeline. We used the dedicated background exposures to perform the pixel-to-pixel background subtraction for each target in stage 2. The stage2 images were then resampled and combined into a final data cube through the Calwebb\_spec3 processing (stage3) using the ‘drizzle’ method with a spaxel size of 0.05$\arcsec$ to enhance the spatial resolution. The resulting data cubes display the well-known sinusoidal modulations in the spectrum caused by the limitation of the pipeline to correctly trace the target across the IFU slices during the rectification process. To correct these artifacts, also known as ``wiggles'', we applied a custom code similar to that developed by \citealt{2023A&A...679A..89P} that masks out the detected lines and models and subtracts the sinusoidal variations (see W.\ Matzko et al.\ 2025, in prep). After completing the pipeline reduction, we removed any remaining pixel outliers from the data cubes using the methods outlined in \citealt{2024PASP..136d4503H}.

\section{Analysis}\label{sec:analysis}

\subsection{Emission line fitting and spatial maps generation}\label{sec:emission}

\begin{figure}[h!]
     \centering
     \includegraphics[width=1\linewidth]{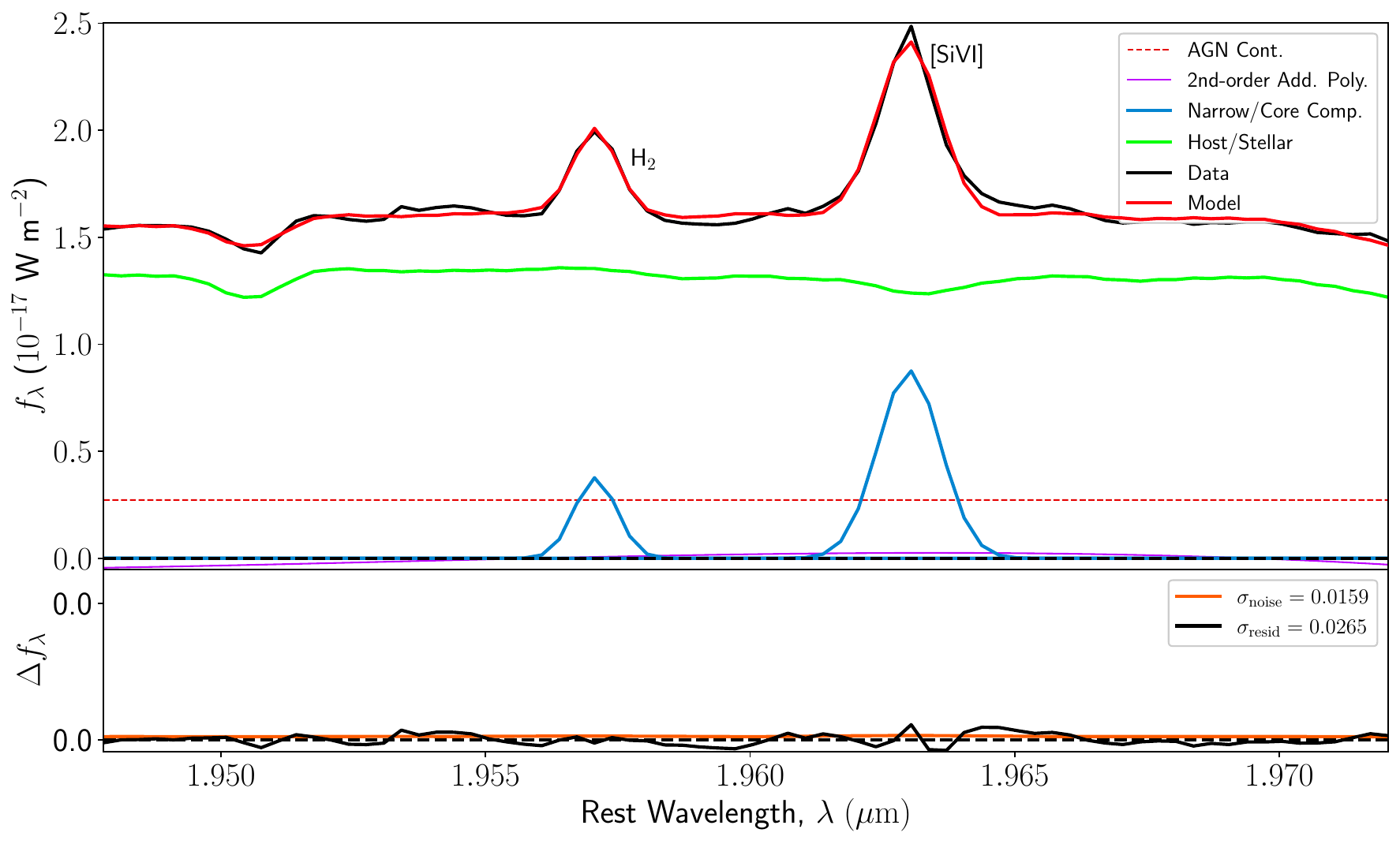}\\ (a)\\
     \includegraphics[width=1\linewidth]{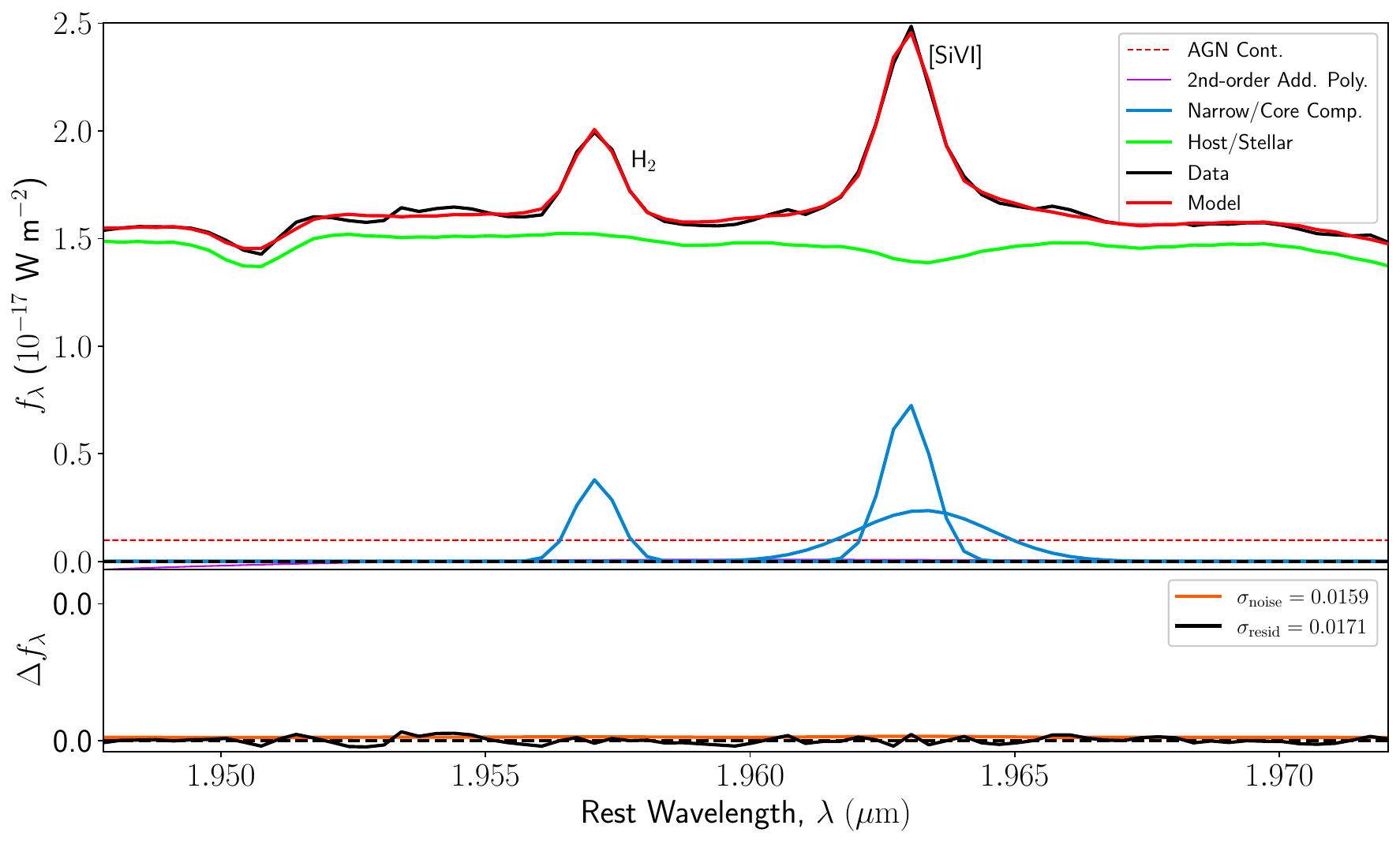}\\ (b)
     \caption{ Examples of (a) one Gaussian component and (b) two Gaussian components fit to the [Si\,VI]  line profile for J0954 for the same single spaxel. In each panel, the black spectrum is the observed data, the solid red line is the best fit, and the blue lines represent the Gaussian components. The residuals after subtraction from the best fit are shown at the bottom of each panel. We see that the single component in (a) does not adequately fit the observed line profile; based on the F-test and CHI2RATIO test, a second component is  justified. A similar second component was also required in the central spaxels of [Si\,VII], [Si\,IX], [Mg\,IV], [Mg\,VIII] and [Ar\,VI] lines in J0954 and J1009.}
     \label{fig:2comp_fit}
\end{figure}

To fit the various coronal emission lines (Table \ref{tab:CL}) and generate spatial flux and velocity maps, we used the Bayesian AGN Decomposition Analysis for Sloan Digital Sky Survey (SDSS) Spectra \citep[BADASS v10.2.0,][]{2021MNRAS.500.2871S} \footnote[1]{\url{https://github.com/remingtonsexton/BADASS3}} code. Through Markov Chain Monte Carlo routines, BADASS performs simultaneous multicomponent fits to emission-line spectra. Individual spaxels in the NIRSpec cubes were fit iteratively across the full spatial extent of the data cubes, where a third-order Legendre polynomial was used for the continuum. The host galaxy and the stellar line-of-sight velocity distribution were fit using the penalized pixel fitting method (pPXF, \citealt{2004PASP..116..138C}) with templates from the eMILES \citep{2016MNRAS.463.3409V} stellar library.  Emission lines were modeled using Gaussian profiles. A secondary Gaussian component to the fit was included based on the line-testing option that is implemented within BADASS. We used a combination of multiple line testing options such as CHI2\_RATIO and a modified F-test. The CHI2\_RATIO test is the ratio of the reduced chi-squared values for each fitted model, and can be interpreted as the fraction of the difference in the residuals weighted by the noise. The modified F-test generates a confidence between 0 and 1 of the Monte-Carlo resampled maximum likelihood values of the simple and complex models. If the confidence is 0.9, the difference between residuals is significant, and an additional component is justified with 90$\%$ confidence. If the value of the CHI2\_RATIO was $>$ 0.1 and if the modified F-test confidence parameter was $>$ 0.9, then adding a second component provided a significant improvement to the fit and was thus justifiable. This secondary component was restricted to have a larger velocity offset from the rest-frame wavelength (see Table \ref{tab:CL}) and a larger dispersion than the primary component (See Figure \ref{fig:2comp_fit} for an example fit to a single spaxel and also Section \ref{sec:wiggles}). Due to the diverse emission profiles in our data, we left the amplitude and width as unrestricted free parameters for the primary component. We also set a signal-to-noise ratio (S/N) threshold of three for all lines and components within each spaxel to determine a detection. Uncertainties were derived from the error spectrum and the random errors associated with the Bayesian fitting process.

\begin{figure*}[ht!]
    \centering
    \includegraphics[width=1\linewidth]{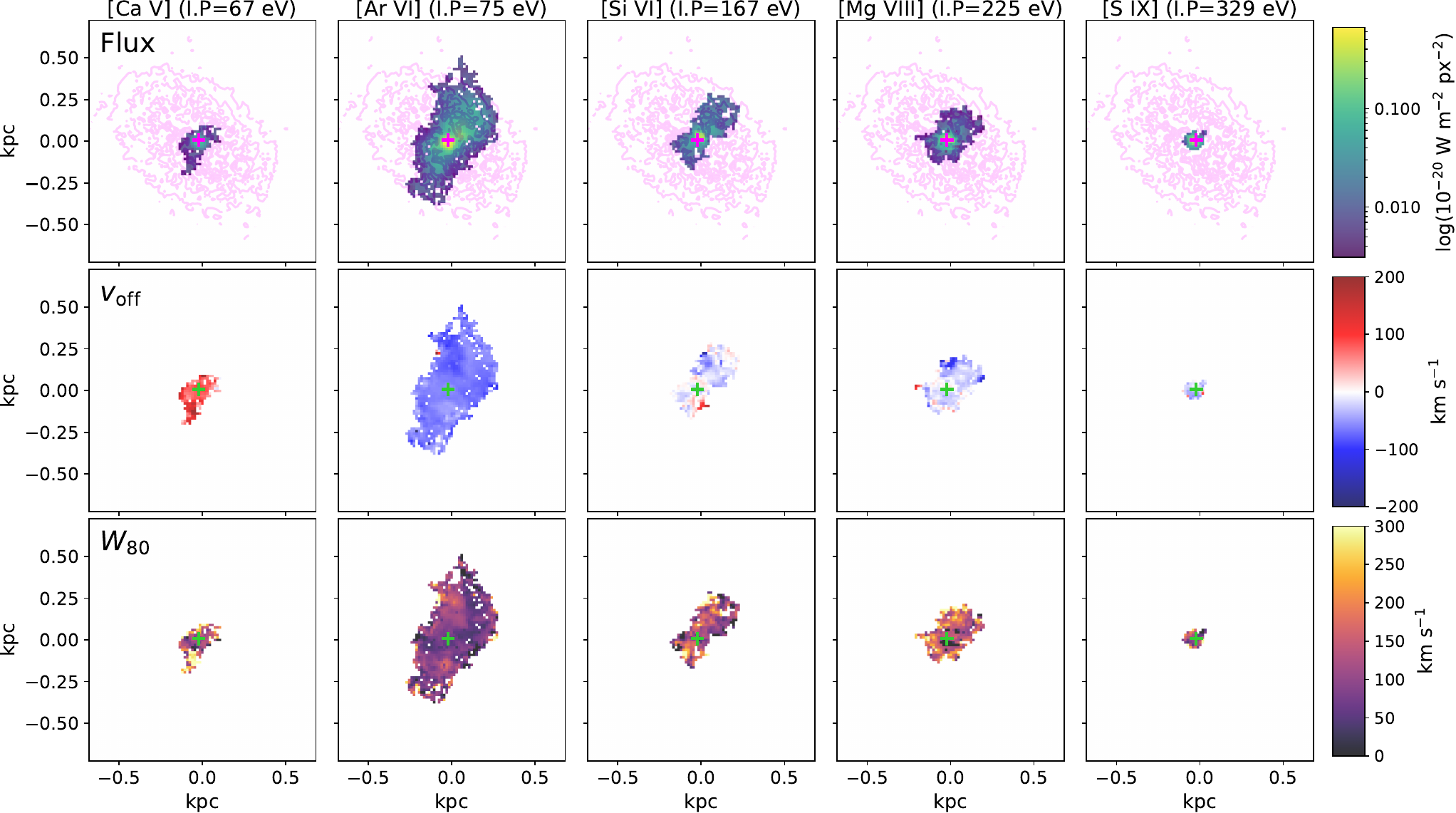}
    \caption{Flux, v$_{\mathrm{off}}$ and W$_{80}$ maps of the narrow component for select CLs in J1009, in order of increasing I.P. The axes represent the physical scale in kpc.  The magenta contours indicate the galaxy continuum, and we find that the CLs tend to be oriented perpendicular to the major axis of the galaxy continuum. The magenta and green crosses indicate the peak of the continuum flux and the potential AGN location (the slight ($\sim$1 pixel) mismatch in the peak can be attributed to PSF wandering between the different NIRSpec filters). The maps are oriented with North pointing up and East to the left.}
    \label{fig:Flux maps J1009}
\end{figure*}

\subsection{Flux, v$_{\mathrm{off}}$ and W$_{80}$ maps}\label{sec:flux_maps}

Maps of the narrow component for a sample of CLs detected in our sample can be seen in Figure \ref{fig:Flux maps J1009}. The v$_{\mathrm{off}}$ shows the offset velocity (calculated from the best-fit Gaussian) from the expected central wavelengths from the systemic redshift (calculated from a combination of stellar absorption features and the strong Paschen $\alpha$ $\lambda$1.873 $\mu$m line) of the associated CL. The central wavelengths of each line were determined from the wavelengths compiled by CLOUDY\footnote[2]{\url{https://linelist.pa.uky.edu/atomic/}}. W$_{80}$ is the line width defined to encompass 80\% of the total flux, such that W$_{80}$ = v$_{90}$ - v$_{10}$, where v$_{10}$ and v$_{90}$ are the velocities at the 10th and 90th percentiles of the total flux, respectively, calculated starting from zero intensity on the red side of the line. 
For a Gaussian, W$_{80}$ is 1.0833 times the Full Width at Half Maximum (FWHM). Thus, W$_{80}$ can be used as an indicator of turbulent or disturbed gas in the galaxy or as the distribution of gas velocities in an outflow \citep{2013MNRAS.436.2576L, 2017A&A...606A..36C, 2020A&ARv..28....2V, 2021MNRAS.505.5469S}. All reported W$_{80}$ values have been corrected for instrumental broadening. The flux maps indicate that most of the emission arises from a compact source at the center of each target. While several previous studies have suggested preferentially blue-shifted emission line profiles for CLs \citep{2009MNRAS.397..172G, 2021ApJ...922..155M}, we do not observe such a systematic preference. On the contrary, there seems to be a variety of observed v$_{\mathrm{off}}$ within the CL, ranging from blue-shifted to red-shifted values. 

\subsection{Presence of a broad component}
\label{sec:wiggles}
\begin{figure*}[ht!]
    \centering
    \includegraphics[width=1\linewidth]{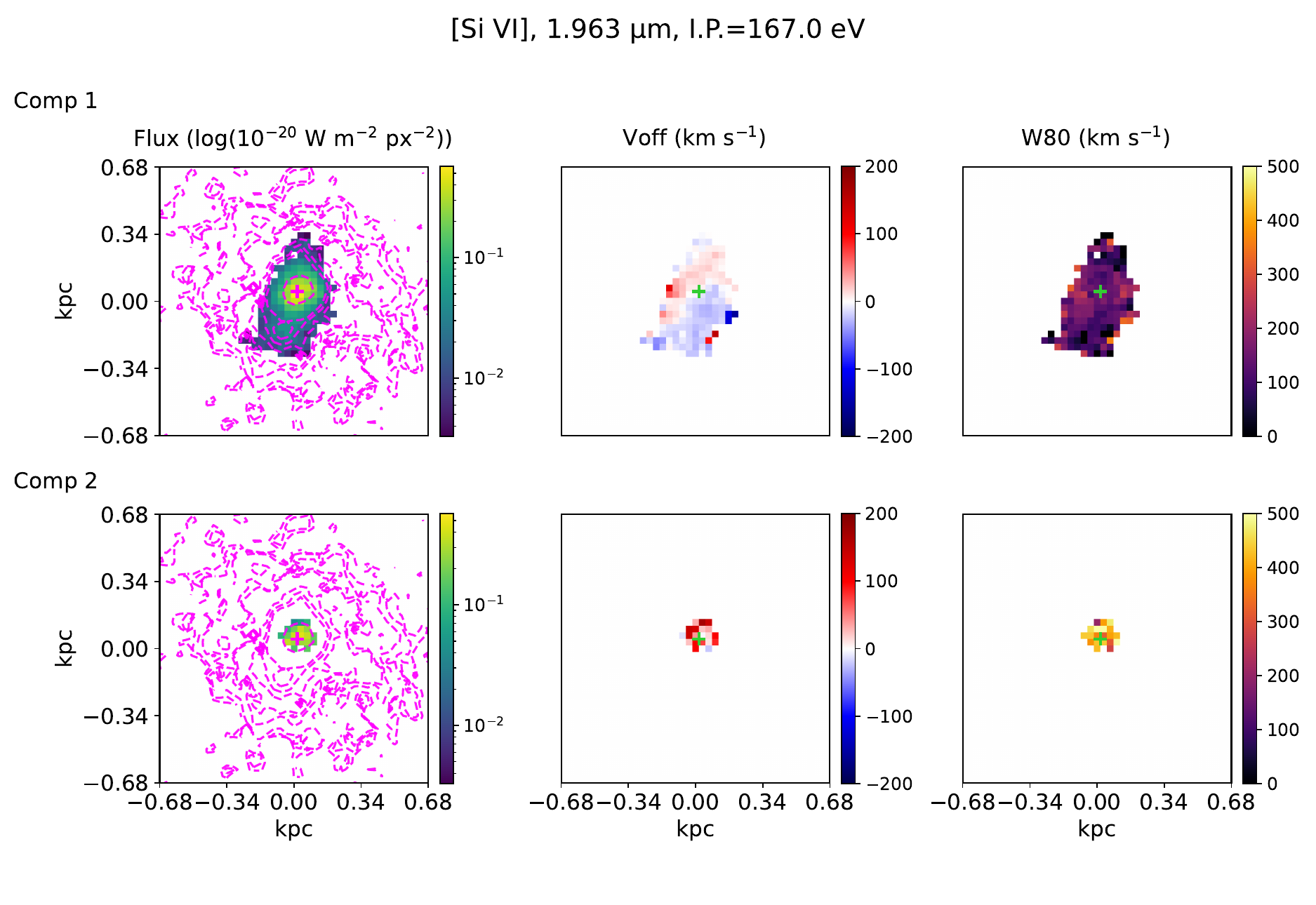}
    \caption{Flux, v$_{\mathrm{off}}$ and W$_{80}$ maps, showing the two components in the [Si\,VI] line in J0954. The magenta contours trace the galaxy continuum and the cross indicates the peak continuum flux corresponding to the AGN position. The detected second component (bottom panel) across all the lines is found to be more compact than the first component, is concentrated in the central region and has larger W$_{80}$ values than the first component.}
    \label{fig:Flux maps J0954}
\end{figure*}

\begin{figure*}[ht!]
  \centering
  \includegraphics[width=\linewidth]{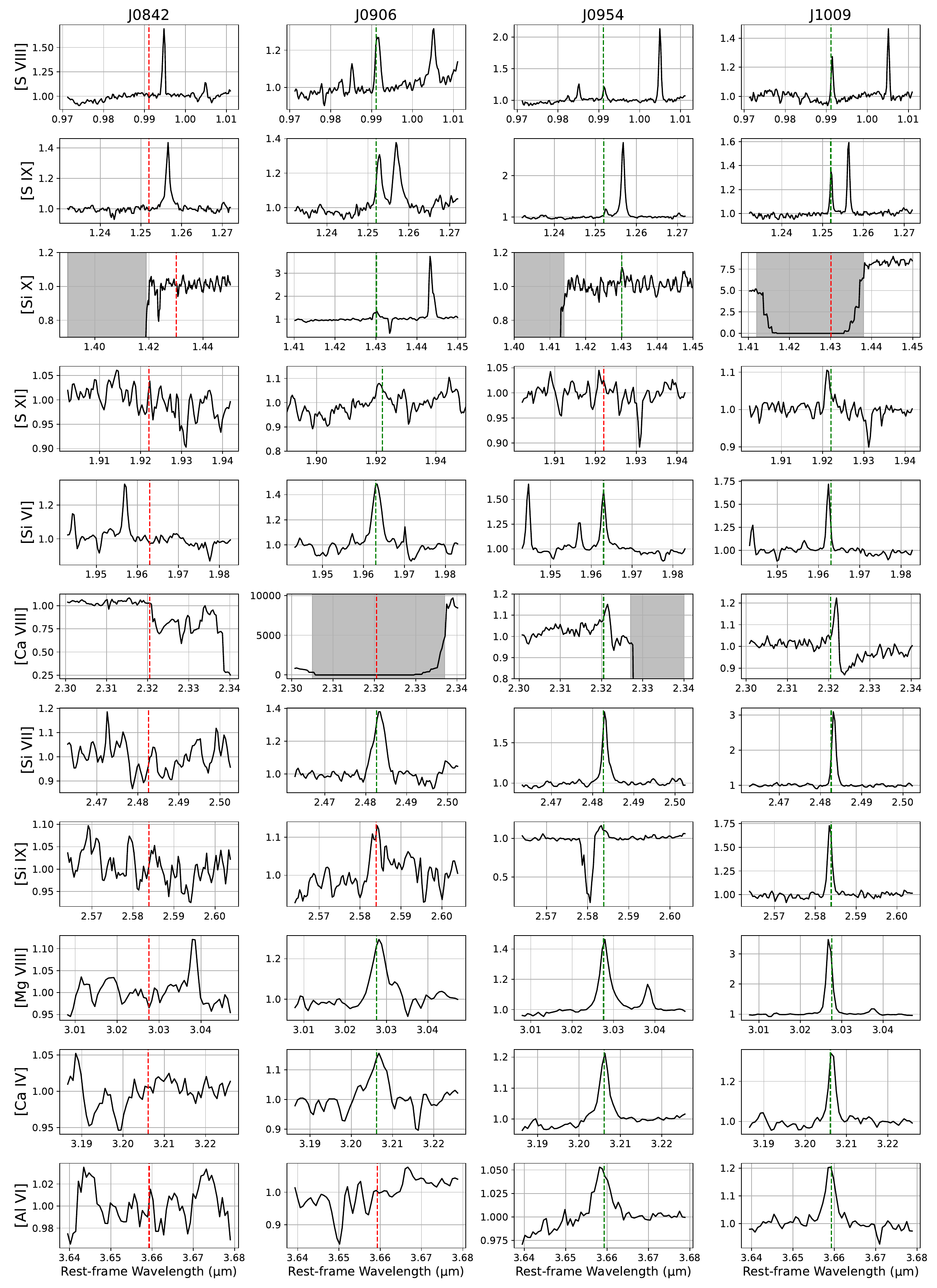}
\end{figure*}
\begin{figure*}[ht!]\ContinuedFloat
  \centering
  \includegraphics[width=\linewidth]{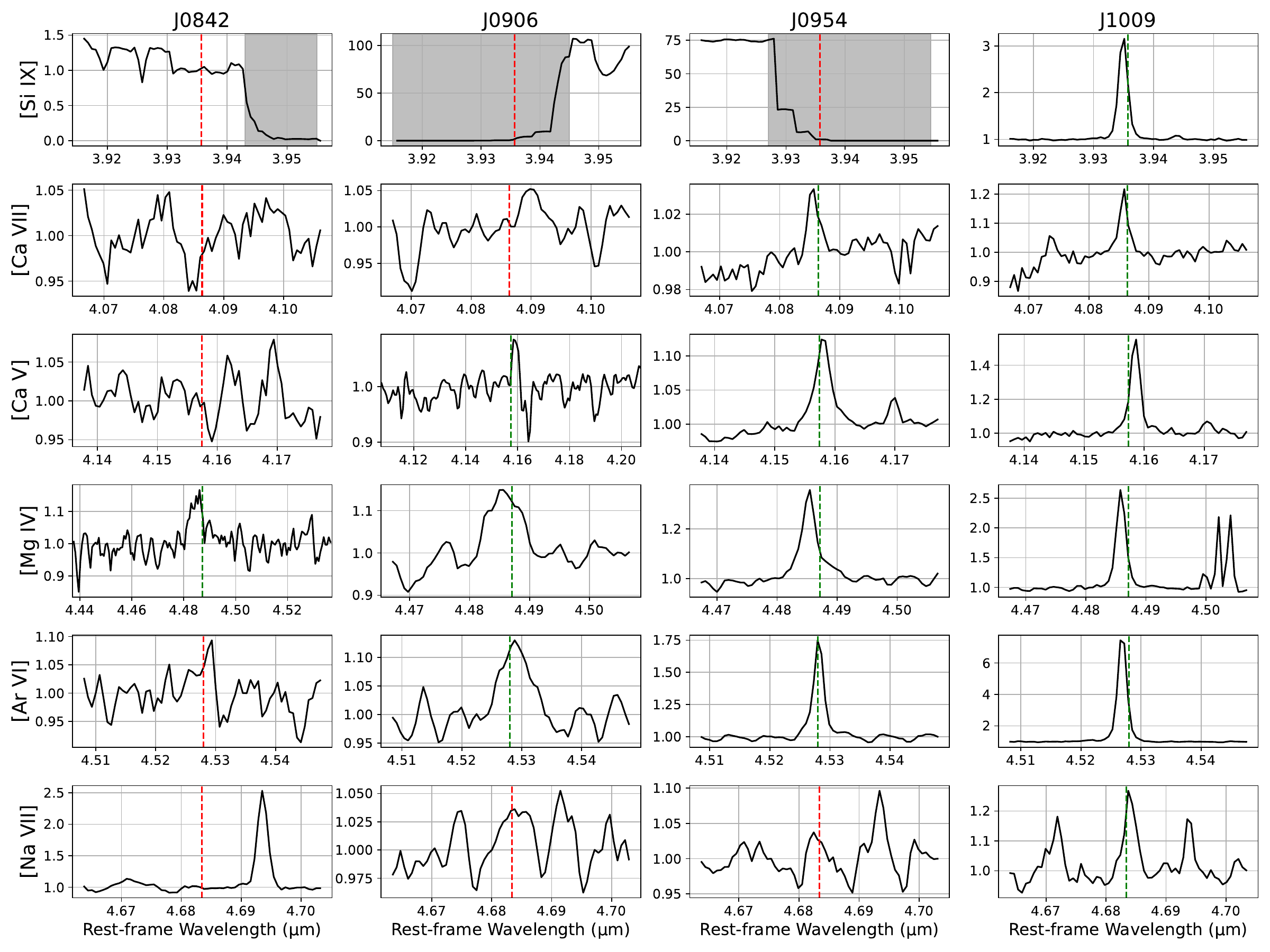}
  \caption{Zoomed-in spectra taken from a 0.3$\arcsec$ radius aperture centered on the galaxy nucleus showing the detected ($>$3$\sigma$; green dotted lines) CLs in the four targets. The non-detections are indicated by the red dotted lines. Although in certain panels, there may be features that are indicative of CL lines, we only claim a detection if the line is $>$3$\sigma$ above the noise. The continuum flux is normalized to unity and the systemic redshift was used to shift the spectra to rest-frame wavelength. The JWST/NIRSpec detector gaps are indicated by the gray shaded region.}
  \label{fig:Spectra}
\end{figure*}

Two of the targets (J0954 and J1009) have a significant number ($\ge$4) of CLs that show signs of having broader wings than a single Gaussian (Figures \ref{fig:2comp_fit} and \ref{fig:Flux maps J0954}), as determined by the line-component testing for the spectra in the individual spaxels (Section \ref{sec:emission}). Previous works have indicated that the optical coronal emission line profiles can be fitted by two-Gaussian components \citep{2005MNRAS.364L..28P, 2006ApJ...653.1098R, 2011ApJ...739...69M}. These studies also find that the first component extends to large distances (0.7 - 3.8 kpc). The second broader, weaker component cannot be measured outside of the bright nucleus ($<$0.03 kpc). We find a similar trend in the two dwarf galaxies where broad CLs are detected (J0954 and J1009). We also find that the broad component is detected only in the stronger lines of [Si\,VI], [Si\,VII], [Mg\,IV], [Mg\,VIII], and [Ar\,VI]. The broad component is too weak to  detect in weaker lines, including those with larger I.P. as well as in Ca and Al lines which have similar I.P. as the aforementioned lines. The second component in all targets is significantly broad (W$_{80}$ $>$ 300 km s$^{-1}$), but does not have significant velocity offset. For one of the targets (J0906), a second component was not included based on the line tests, but we find that the narrow component is significantly broad (W$_{80}$ $>$ 300 km s$^{-1}$). 

The presence of continuum ``wiggles'' (Section \ref{sec:Observations and Data reduction}) in the data significantly limits the detailed characterization of the broad component. This posed challenges in fitting and getting resolved maps of the broad component. To fit the individual spaxels, we had to model the wiggles and remove them from the data. We found that it was impossible to distinguish between the wiggles and the faint wings of the outflows. To quantify the difference between the wiggled and de-wiggled cubes, we extract and fit spatially integrated spectra from each cube using the same aperture. We find that the parameter that was most affected by the uncertainty introduced by the wiggles is the width of the broad component, whereas fluxes and velocity offsets do not change significantly. We find that the width parameters, such as W$_{80}$ and FWHM of the broad components, were lower by 20-30$\%$ in the de-wiggled cube as compared to the wiggled cubes. Thus all the estimates of the W$_{80}$ of the broad component in this work should be treated as lower limits.

\subsection{Determining the extent of the CLs}\label{sec:Determining the extent of the CL}

To measure the spatial extent of each NIR CL, we first identify the central position ($x_{\mathrm{center}}, y_{\mathrm{center}}$) as the pixel with the peak flux. We then locate the furthest spaxel ($x_{\mathrm{ext}}, y_{\mathrm{ext}}$) from the center with a S/N of at least 3. To account for projection effects, we de-project this distance using the galaxy inclination angle i, estimated from the ratio of the semi-minor axis to the semi-major axis, b/a. We compute cos i = b/a, where b/a is the exponential fit b/a value (\texttt{expAB$\_$r}) obtained from the SDSS \texttt{PhotObj} catalog.

The de-projected extent in spaxel units is calculated using Equation (1) from \citealt{2021ApJ...920...62N}:
\begin{equation}
\mathrm{Extent} = \sqrt{(x_{\mathrm{ext}} - x_{\mathrm{center}})^2 + \left[(y_{\mathrm{ext}} - y_{\mathrm{center}}) \cdot \cos i\right]^2}
\end{equation}

We then convert the extent to physical units (kpc) using the \texttt{astropy.cosmology} Python package. We also estimate the extent using alternative definitions, including the half-light radius, 95th-percentile radius (R95) and root-mean-square (rms) radius, and find that the trends described in Section~\ref{ref:CLR} remain valid across these methods.

We then checked whether the CL extents are set by the line strength. First, we varied the S/N threshold from 1.5 to 3$\sigma$ and required the inferred extent to change by $\le$15\%. For non-detections, we computed a 3$\sigma$ PSF-aperture flux limit ($\sigma$ = F/(S/N)) and compared it to the flux predicted from a reference line of the same element ([Si\,VII] for Si, [Mg\,VIII] for Mg, [S\,VIII] for S, and [Ca\,V] for Ca) assuming lines from the same species share a common spatial profile. For elements lacking multiple ionization stages (e.g., [Al\,VI], [Na\,VII]), we compared to the brightest available CL, [Ar\,VI]. We classified a line as absent only when the predicted flux exceeds the 3$\sigma$ limit. Thus, the reported extents trace real emission boundaries rather than sensitivity limits or single-pixel artifacts, within the stated uncertainties.

\section{Results}\label{sec:results}

\subsection{Coronal line detections} \label{sec:coronal line detections}

\begin{deluxetable*}{lcccccccc}[ht!]
\tablecolumns{8}
\tablecaption{Summary of Coronal Line (CL) Emission in the Targets \label{tab:CL}}
\tablehead{
  \colhead{CL} &
  \colhead{$\lambda$ ($\mu$m)} &
  \colhead{I.P. (eV)} &
  \colhead{log $n_{\rm crit}$ (cm$^{-3}$)} &
  \colhead{J0842} &
  \colhead{J0906} &
  \colhead{J0954} &
  \colhead{J1009}
\\
  \colhead{} &
  \colhead{($\mu$m)} &
  \colhead{(eV)} &
  \colhead{(cm$^{-3}$)} &
  \colhead{} &
  \colhead{} &
  \colhead{} &
  \colhead{}\\
\colhead{(1)} & \colhead{(2)} & \colhead{(3)} & \colhead{(4)} & \colhead{(5)} & \colhead{(6)} & \colhead{(7)} & \colhead{(8)}
}
\startdata
$[$S\,VIII$]$   & 0.9911 & 281 & 9.47 & ... & \checkmark$^{\dagger}$ & \checkmark$^{\dagger}$ & \checkmark$^{\dagger}$ \\
$[$S\,IX$]$     & 1.2520 & 329 & 9.42 & ... & \checkmark$^{\dagger}$ & \checkmark$^{\dagger}$ & \checkmark$^{\dagger}$ \\
$[$Si\,X$]$     & 1.4301 & 351 & 6.70 & ... & \checkmark$^{\dagger}$ & \checkmark             & ... \\
$[$S\,XI$]$     & 1.9220 & 447 & 7.98 & ... & \checkmark$^{*}$                   & ...                    & \checkmark \\
$[$Si\,VI$]$    & 1.9630 & 167 & 8.72 & ... & \checkmark$^{\dagger}$ & \checkmark$^{\dagger}$ & \checkmark$^{\dagger}$ \\
\textbf{(Broad)   }      &        &     &      & ... & ...                    & \underline{\boldcheckmark}          & \underline{\boldcheckmark} \\
$[$Ca\,VIII$]$  & 2.3205 & 127 & 6.66 & $^{\dagger}$ & $^{\dagger}$ & \checkmark$^{*}$ & \checkmark$^{\dagger}$ \\
$[$Si\,VII$]$   & 2.4826 & 205 & 8.48 & ... & \checkmark             & \checkmark             & \checkmark \\
\textbf{(Broad)  }       &        &     &      & ... & ...                    & \underline{\boldcheckmark}          & \underline{\boldcheckmark} \\
$[$Si\,IX$]$    & 2.5839 & 304 & 8.04 & ... & ...                    & \checkmark             & \checkmark \\
$[$Mg\,VIII$]$  & 3.0276 & 225 & 6.84 & ... & \checkmark             & \checkmark             & \checkmark \\
\textbf{(Broad) }        &        &     &      & ... & ...                    & \underline{\boldcheckmark}          & \underline{\boldcheckmark} \\
$[$Ca\,IV$]$    & 3.2061 &  51 & 7.10 & ... & \checkmark$^{*}$       & \checkmark             & \checkmark \\
$[$Al\,VI$]$    & 3.6593 & 154 & 5.88 & ... & ...                    & \checkmark             & \checkmark \\
$[$Si\,IX$]$    & 3.9357$^{\#}$ & 304 & 8.04 & ... & ...               & ...                    & \checkmark \\
\textbf{(Broad) }        &        &     &      & ... & ...                    & ...                    & \underline{\boldcheckmark} \\
$[$Ca\,VII$]$   & 4.0864 & 109 & 7.28 & ... & ...                    & \checkmark             & \checkmark \\
$[$Ca\,V$]$     & 4.1574 &  67 & 6.53 & ... & \checkmark$^{*}$       & \checkmark             & \checkmark \\
$[$Mg\,IV$]$    & 4.4871 &  80 & 7.24 & \checkmark & \checkmark$^{*}$  & \checkmark             & \checkmark \\
\textbf{(Broad) }        &        &     &      & ... & ...                    & ...                    & \underline{\boldcheckmark} \\
$[$Ar\,VI$]$    & 4.5280 &  75 & 7.33 & ... & \checkmark$^{*}$       & \checkmark             & \checkmark \\
\textbf{(Broad)}         &        &     &      & ... & ...                    & \underline{\boldcheckmark}          & \underline{\boldcheckmark} \\
$[$Na\,VII$]$   & 4.6834 & 172 & 6.66 & ... & ...                    & ...                    & \checkmark \\
\enddata
\tablecomments{
Columns: (1) Coronal line species; (2) Rest wavelength from \textsc{CLOUDY} ($\mu$m). $^{\#}$ indicates wavelength from the CAFE \citep{2025ascl.soft01001D} linelist\footnote[1]{\url{https://github.com/GOALS-survey/CAFE}}.  
(3) Ionization potential (eV) from the Chianti database\footnote[2]{\url{https://www.chiantidatabase.org}}.  
(4) Critical density (log cm$^{-3}$) from \textsc{CLOUDY}.  (5–8) Detection status in each galaxy: $\checkmark$ = detected with JWST/NIRSpec; \underline{$\boldcheckmark$} = broad component detected; $^{\dagger}$ = previously detected with Keck/NIRES; $^{*}$ = detected only in the 0.3$\arcsec$ integrated spectrum (not across individual spaxels).}
\end{deluxetable*}

\begin{figure*}[ht!]
    \centering
    \includegraphics[width=1\linewidth]{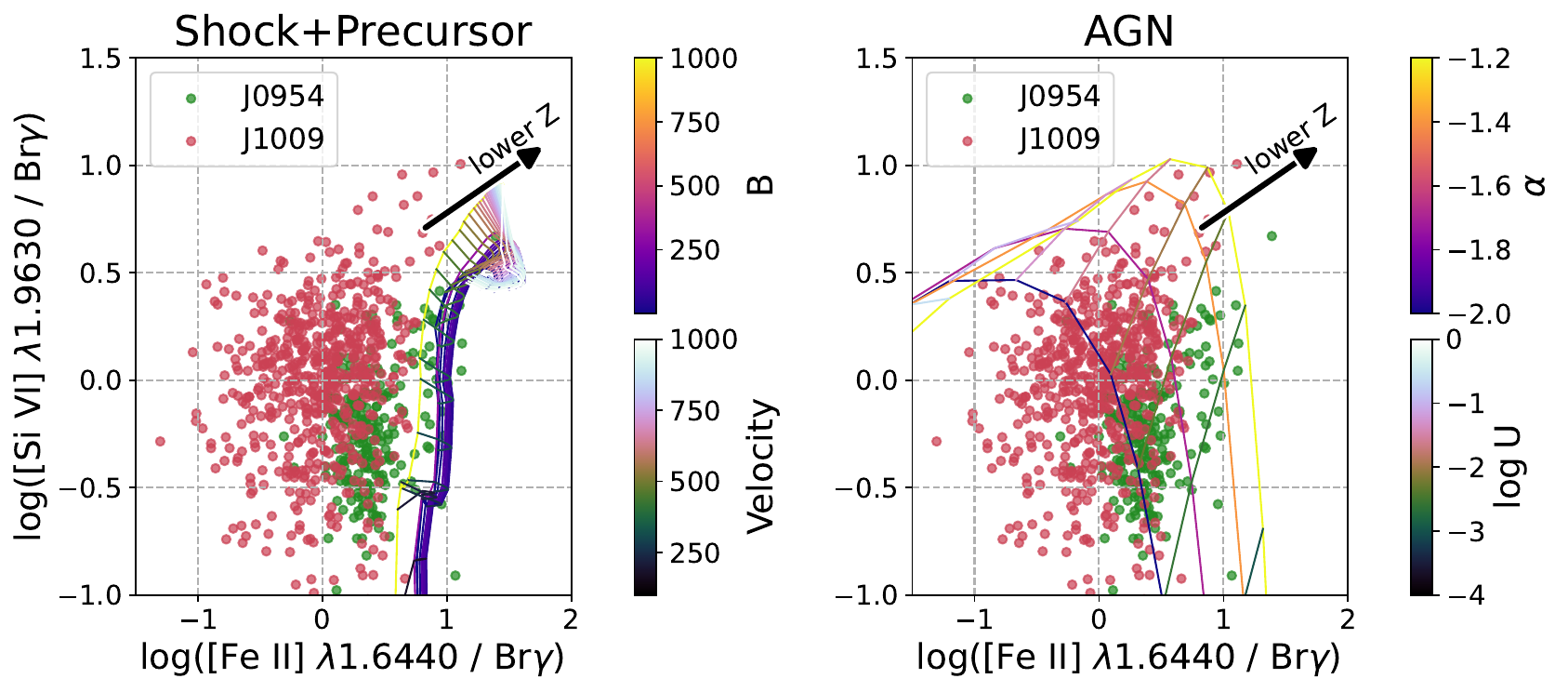}
    \caption{Shock+Precursor model (left) and AGN models (right) for [Si\,VI] for J0954 (green) and J1009 (red), assuming solar metallicities. The free parameters for the Shock+Precursor model (left) are B and velocity, where B is the transverse magnetic field strength in $\mu$G, velocity is the shock velocity in km s$^{-1}$. The free parameters for the AGN model (right) are $\alpha$, which is the power-law index indicating the steepness of the power-law slope and log U, which is the ionization parameter. The points represent values from individual spaxels wherever all the four lines used in the ratios were detected. We find that the observed line ratios largely agree with the AGN photoionization model for these two targets. The black arrow indicates the direction of the shift in the spaxels for lower metallicities. The shift is found to be $\sim$0.1 dex at 0.5\,Z$_\odot$ and $\sim$0.2 dex at 0.25\,Z$_\odot$. }
   \label{fig:Shock+agn models both}
\end{figure*}

Previous ground based NIR observations using Keck/NIRES \citep{2021ApJ...911...70B} revealed the presence of high-ionization CLs in the four targets between 0.9 - 2.3 $\mu$m. We now report the detection of new CLs due to the improved sensitivity as well as in the extended wavelength range provided by JWST/NIRSpec (0.9 - 5 $\mu$m). \citealt{2021ApJ...911...70B} reported a single CL, [Ca\,VIII], in J0842, which we do not detect in our JWST/NIRSpec observations. However, the expected wavelength of [Ca\,VIII] lies very close to the detector gap in the data cube, which makes it challenging to detect robustly. We also detect [Mg\,IV] at 4.48 $\mu$m, which is the only coronal line reliably observed in this target with NIRSpec. For J0906, \citealt{2021ApJ...911...70B} identified five CLs from ground-based data. We recover four of these in the NIRSpec spectra. [Ca\,VIII] is not detected as it falls within the transmission gap between the two NIRSpec detectors. In addition to the previously known lines, we detect six new CLs, increasing the total number of NIR-detected CLs in this galaxy to 11. In J0954, three NIR CLs had been previously reported \citep{2021ApJ...911...70B}. Our observations reveal 11 additional CLs, bringing the total to 14. Finally, J1009 exhibits the richest coronal line spectrum among the four targets. While previous ground-based observations identified four CLs, our NIRSpec data reveal 12 additional lines. These include several highly ionized species such as [S\,XI] (I.P. = 447 eV), as well as multiple ionization states of Ca and Na. For a complete list of detected CLs in the four targets see Tab. \ref{tab:CL} and Figure \ref{fig:Spectra}.

Overall, we detect all CLs reported by \citealt{2021ApJ...911...70B} except those that fall in the JWST detector gap. We do not confirm the [Ca\,VIII] detection reported by \citealt{2021ApJ...911...70B}; given its 2.2$\sigma$ significance, it may have been a noise artifact. Among the new lines, only one ([S\,XI]) lies within the Keck/NIRES wavelength range but was likely missed due to lower sensitivity or imperfect atmospheric subtraction. [Si\,X] was not detected with Keck/NIRES in J0954 but is seen with NIRSpec, likely due to the higher spectral resolution. The remaining new detections are enabled by the extended wavelength coverage of NIRSpec.

Three of the four targets also show optical CLs in prior SDSS, Keck/LRIS, Keck/KCWI and Gemini/GMOS spectra \citep{2019ApJ...884...54M, 2021ApJ...911...70B}. J0906 has [Ne\,V] $\lambda$3426, [Fe\,X] $\lambda$6374 and [Fe\,VII] $\lambda$$\lambda$$\lambda$5721, 5159, 6087. J0954 has [Ne\,V] $\lambda$$\lambda$ 3346, 3426, [Fe\,V] $\lambda$3839, and [Fe\,VII]$\lambda$$\lambda$ 5159, 6087. J1009 has possible detections of [Fe\,VII] $\lambda$$\lambda$5721, 6087. J0842 has no previously detected optical CLs.

The CL peaks in all targets coincide with the K-band continuum peak, and, within each target, the CL peaks coincide with one another, strongly suggesting that they are produced by the central AGN source. Although CLs have also been attributed to tidal disruption events (TDEs; \citealt{2023MNRAS.519.2035H, 2024MNRAS.528.7076C}), a comparison between the JWST spectra and Keck/NIRES observations taken $\sim$5 years earlier \citep{2021ApJ...911...70B} shows no evidence of fading of the CLs. A five-year interval is consistent with the expected fading timescales of TDE-driven CLs \citep{2013ApJ...774...46Y}. Thus, it is unlikely that these CLs are produced by a TDE and are instead directly ionized by the central AGN.

\begin{figure*}
    \centering
    \includegraphics[width=1\linewidth]{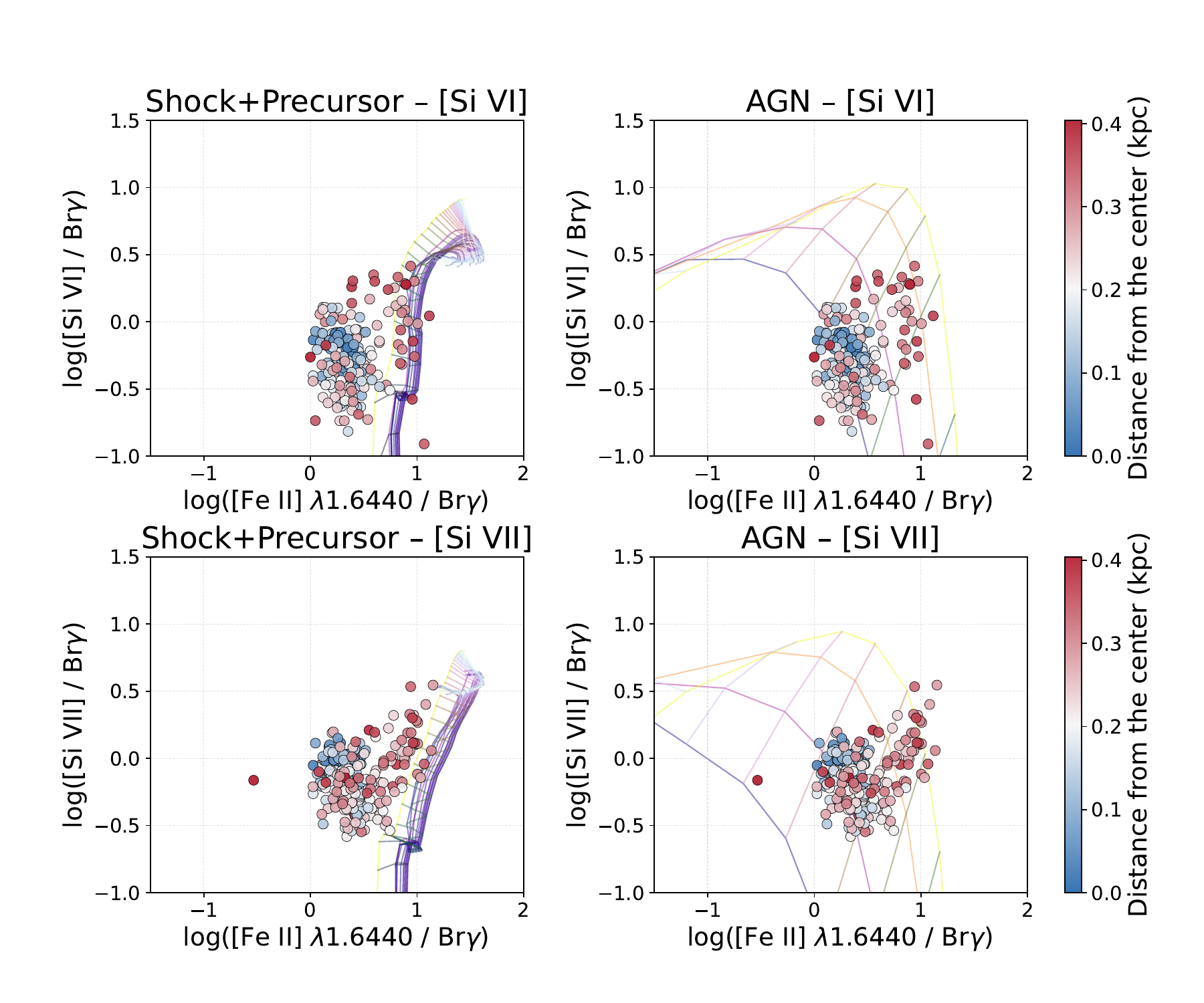}
    \caption{Shock+Precursor model (left) and AGN models (right) for [Si\,VI] and [Si\,VII] in J0954. The grid parameters follow the same scheme as in Figure \ref{fig:Shock+agn models both} and the color bar indicates the distance from the center in pixels. The ratios in J0954 remain broadly consistent with AGN photoionization even up to 0.4 kpc from the nucleus, with only minor overlap with shock+precursor models.}
    \label{Shock models with distance}
\end{figure*}

\subsection{Source of Ionization}\label{sec:Source of ionizing mechanism}
While high-ionization CLs are generally attributed to photoionization by the hard radiation field of the AGN, some studies have proposed that they may also be produced by fast radiative shocks associated with outflows \citep{1995ApJ...455..468D}. Given that the dwarf galaxies in our sample exhibit evidence of outflows, we investigate the dominant ionization mechanism responsible for the observed CLs.

To disentangle the contributions from AGN photoionization and shocks, we utilize diagnostic flux ratios such as [Si\,VI]/Br$\gamma$ and [Si\,VII]/Br$\gamma$ vs. [Fe II] 1.64 $\mu$m/Br$\gamma$. The [Fe\,II] line is a well-established tracer of shock excitation \citep{2013ApJ...775..115U}, whereas Br$\gamma$ traces star formation and the overall ionizing photon field. These ratios are compared against both AGN photoionization models from \citealt{2004ApJS..153...75G,2004ApJS..153....9G} and shock+precursor models from \citealt{2008ApJS..178...20A}, as implemented in the IDL Tool for Emission-line Ratio Analysis (ITERA; \citealt{2010NewA...15..614G}). We plot the ratios obtained from the total fluxes of the emission lines from individual spaxels in two of the galaxies (J0954 and J1009), where all the lines are present in Figure \ref{fig:Shock+agn models both}.

The shock models account for both the shocked gas and the photoionized precursor region ahead of the shock front. The free parameters include the shock velocity ($v_{\rm shock} = 10$–$1000$ km s$^{-1}$) and the magnetic parameter $B/n^{1/2}$ (ranging from $10^{-4}$ to $10,\mu$G cm$^{3/2}$), where $B$ is the transverse magnetic field strength. The AGN models assume a power-law ionizing spectrum ($F_{\nu} \propto \nu^{\alpha}$) with $5$ eV $< \nu <$ 1000 eV, with a range of possible slopes $\alpha$, and a range of ionization parameters $U = n_{\rm ion}/n_e$, where $n_{\rm ion}$ is the ionizing photon density and $n_e$ is the electron density. 

As shown in Figure \ref{fig:Shock+agn models both}, most of the spatially resolved emission-line ratios for J0954 and J1009 align more closely with the AGN photoionization models. Within the AGN photoionization grids, the points scatter around the gridline with more negative $\alpha$, indicating a steeper power-law spectrum, which can be linked to factors like higher accretion rates and potentially lower black hole masses \citep{2005A&A...432...15P, 2010A&A...512A..58I}. The results are still valid when we use other CLs (i.e. [Si\,VII]). We assume solar metallicity for the grids, similar to \citealt{2020ApJ...905..166L} and \citealt{ 2021ApJ...911...70B}, but we find that if we assume lower metallicities for these dwarf galaxies \citep{2020ApJ...903...58C}, the results are still broadly valid. There is an upward shift of the spaxels towards the shock+precursor grids as well as towards higher $\alpha$ and log U values in the AGN grids for lower metallicities, but this shift is 0.1 dex for 0.5\,Z$_\odot$ and 0.2 dex for 0.25\,Z$_\odot$.

\begin{figure*}[ht!]
    \centering
    \includegraphics[width=1\linewidth]{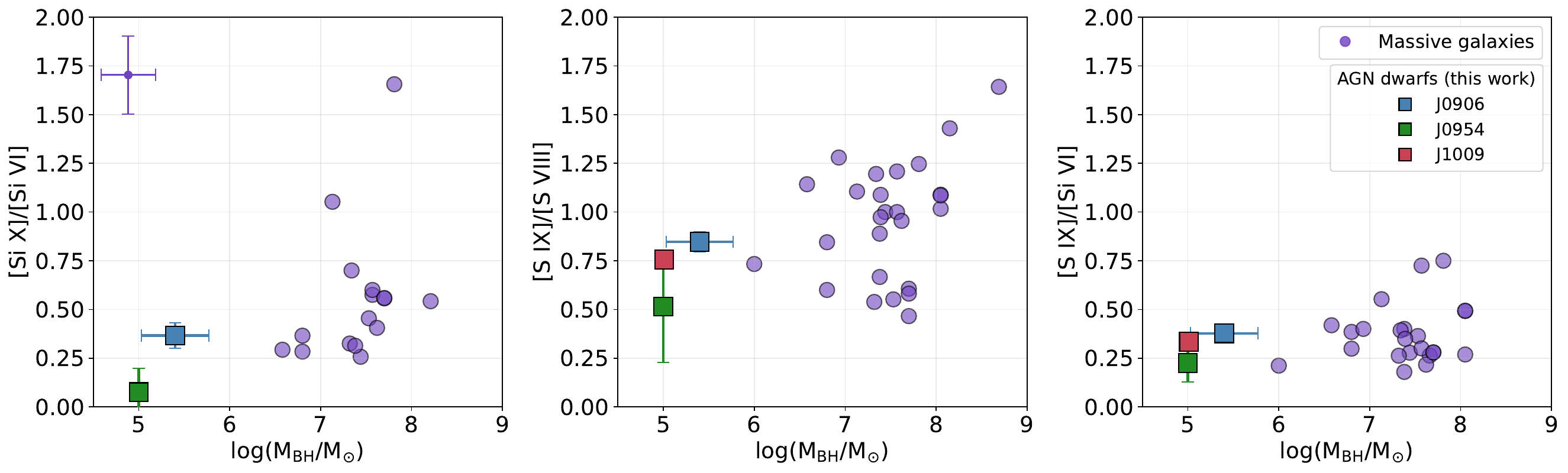}
    \caption{Flux ratios of commonly observed NIR CLs from this work, compared with measurements in massive galaxies from \citealt{2006A&A...457...61R, 2011ApJ...739...69M, 2011ApJ...743..100R, 2017MNRAS.467..540L, 2024ApJ...965..103B}. Black hole masses for the dwarf AGN are taken from the references listed in Table~\ref{tab:tab1}. We adopt a uniform uncertainty of 0.5 dex on the black hole masses, based on the average of the quoted errors where available; this is illustrated by the purple error bar in the top left of the first panel. The flux ratios of lines within the same species, [Si\,X]/[Si\,VI] (first panel) and [S\,IX]/[S\,VIII] (second panel), in dwarf galaxies lie toward the lower end of the distribution compared to those in more massive systems.}
    \label{fig:ratios}
\end{figure*}

In J0954, a larger fraction of the flux ratios overlap with the shock+precursor model predictions. This fraction likely goes up if we consider lower metallicities. To further examine this, we plot the line ratios as a function of radial distance from the nucleus. We find that near the galaxy center the ratios are consistent with pure AGN photoionization, even while assuming lower metallicities. At larger radii, they shift towards values predicted by the shock+precursor models (Figure \ref{Shock models with distance}). This spatial trend suggests that, while AGN photoionization dominates near the nucleus, shocks driven by outflows may contribute to line excitation at larger distances from the central source, potentially accounting for the extended nature of some CLs. These findings are consistent with previous studies in massive galaxies that proposed a composite ionization scenario involving both AGN and shocks \citep{2006ApJ...653.1098R}. Furthermore, since the CL emission follows the AGN ionization bicone (see Section \ref{ionized}), declines with radius, and does not show excess emission at larger distances, any contribution from shocks is likely small.

\section{Discussion}\label{sec:discussion}
\subsection{Non-Uniform Frequency of the Coronal Lines}

Previous studies of CLs in dwarf galaxies have found that few systems display all CLs detectable in a given spectral window simultaneously \citep{2006A&A...457...61R}. The most frequent CLs detections are from Si and S, with the lower-ionization lines (I.P. $<$ 300 eV) typically being stronger. Lines from other elements such as Ca, Al, and Fe are generally weaker and less often detected. Ca and Al can be suppressed by metallicity and depletion onto dust, while Fe may be less affected. Other CLs, such as those from Mg, have rarely been studied because their wavelengths were largely inaccessible in the pre-JWST era. In massive galaxies, it was initially assumed that the frequency and strength of CLs should depend on the orientation of the torus with respect to the line of sight \citep{1998ApJ...497L...9M, 2000AJ....119.2605N, 2009MNRAS.397..172G}, but subsequent work has found little to no difference in the number of detected CLs between Type 1 and Type 2 AGN \citep{2006A&A...457...61R, 2010MNRAS.405.1315M, 2011ApJ...743..100R}. The absence of CLs above certain ionization energies may instead point to a softer ionizing continuum. In some cases, higher-I.P. lines are detected while lower-I.P. lines of different species are absent \citep{2017MNRAS.467..540L, 2024ApJ...975...60A}, which can be explained by differences in elemental abundances or lack of sensitivity from ground-based NIR instruments. In our sample, we mostly find that, within a given species, there are very few cases where a higher-I.P. line is detected but the corresponding lower-I.P. line is not.

To investigate the origin of the non-uniformity in CL detections, we first examine diagnostics of the ionizing source. J0842 is the only target that falls in the composite region of the BPT diagram when using Keck/LRIS fluxes from \citealt{2019ApJ...884...54M}. It may either lack a powerful, hard-spectrum AGN relative to the others, or its AGN signatures may be significantly diluted by stellar contamination. All other targets (J0906, J0954, J1009) lie in the Seyfert region of the BPT diagram \citep{2019ApJ...884...54M}. Both J0906 and J0954 have strong X-ray detections consistent with AGN activity \citep{2017ApJ...836...20B}, and J0906 additionally hosts a parsec-scale radio jet \citep{2020MNRAS.495L..71Y}. J0906 also has the most luminous CLs in our sample, and the median FWHM of its narrow CL component exceeds the median FWHM of both the narrow and broad components in the other three targets, consistent with a powerful central engine and energetic nuclear gas.

J1009 exhibits the richest coronal-line spectrum, with 16 lines detected. As shown in Section \ref{sec:Source of ionizing mechanism}, its log $U$ values are less negative than those of J0954, likely contributing to the larger number of CLs, including [S\,XI] (I.P. = 447 eV), which is not detected in J0954. J1009 also has the lowest stellar mass among our targets and, based on black hole–stellar mass relations \citep{2015ApJ...813...82R}, is expected to host a relatively low-mass black hole. Previous studies have shown that lower-mass black holes are more likely to exhibit multiple strong high-ionization CLs \citep{2018ApJ...861..142C}. As black hole mass decreases, the Schwarzschild radius decreases and the accretion disk temperature increases, shifting the ionizing continuum to higher energies and enhancing the production of high-ionization lines. The CL detections in J1009 are consistent with this scenario.

\begin{table}[h!]
    \centering
    \begin{tabular}{cccc}
       \hline
       \hline
        Target & Pa$\alpha$/Pa$\beta$ & E(B-V) & E(B-V) (from H$\alpha$/H$\beta$)\\
        \hline
        J0842 & 2.11 & 0.00 & 0.00\\
        J0906 & 2.51 & 0.44 & 0.00\\
        J0954 & 2.48 & 0.41 & 0.049\\
        J1009 & 2.28 & 0.18 & 0.166\\
        \hline
    \end{tabular}
    \caption{Pa$\alpha$/Pa$\beta$ ratios and extinction values [E(B‒V)] for the four targets, derived from the narrow component fluxes from the spectra extracted using a 0.3$\arcsec$ central aperture radius. The extinctions were calculated assuming a Cardelli law with R$_V$ = 3.1 and an intrinsic Pa$\alpha$/Pa$\beta$ ratio of 2.133, which is typically assumed for HII regions. Thus, these E(B-V) estimates are uncertain, and do not always agree with extinctions derived assuming an intrinsic ratio of H$\alpha$/H$\beta$= 3.1 \citep{2021ApJ...911...70B}, shown for comparison.}
    \label{tab:extinction}
\end{table}

\begin{figure*}[ht!]
    \centering
    \includegraphics[width=1\linewidth]{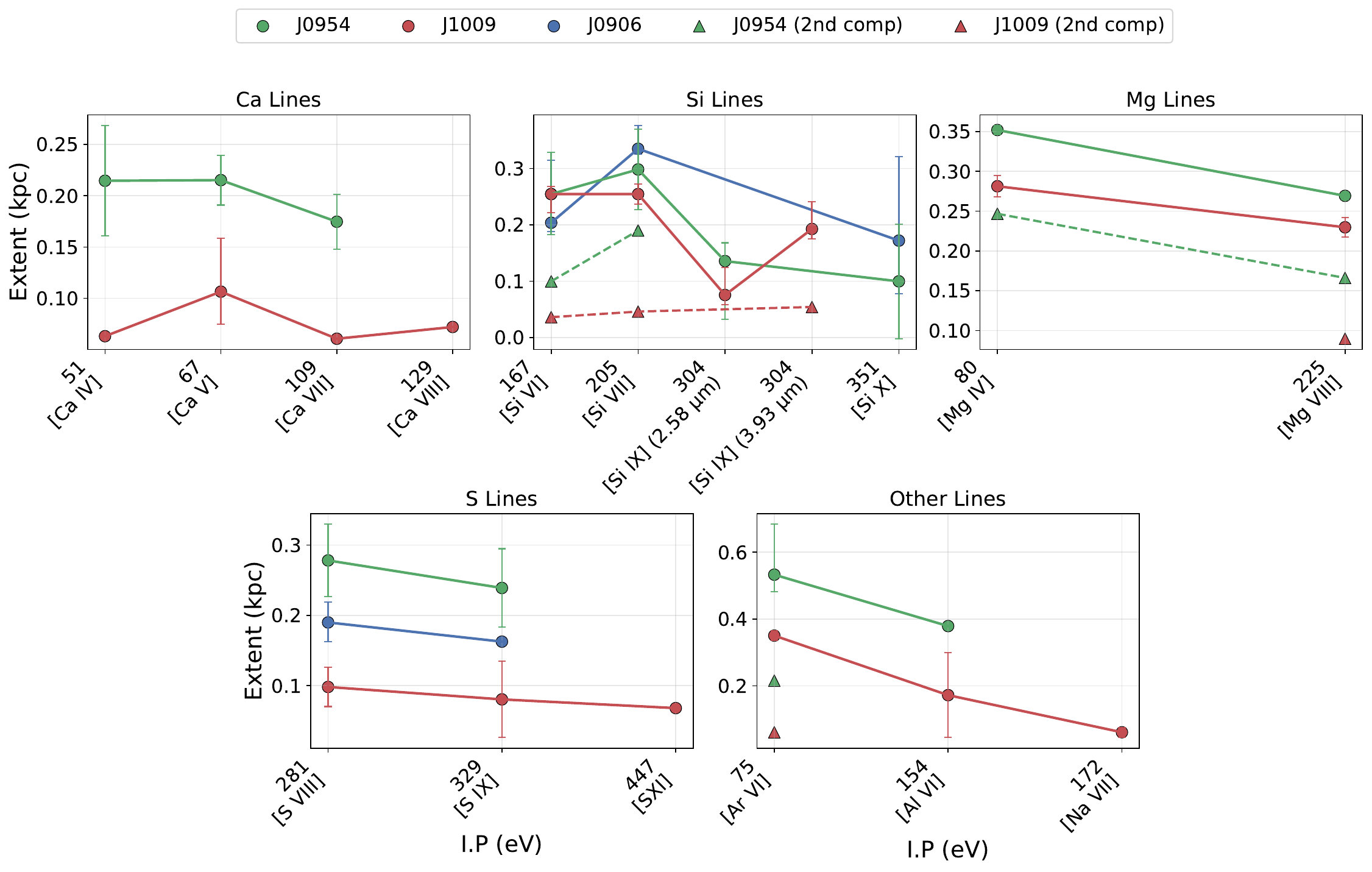}
    \caption{Extents of the various CLs (determined as the radius from the center using the methods outlined in Section \ref{sec:Determining the extent of the CL}), separated by species, arranged in increasing order of ionization potential on the x-axis. Error bars show asymmetric uncertainties on the CL extents derived from the range of values obtained when varying the S/N threshold used to define significant emission. The solid line indicates the primary component and the dashed line indicates the secondary component, wherever detected. We find that lines with the highest ionization potentials have smaller extents.}
    \label{fig:Extents}
\end{figure*}

We next consider the role of dust. Balmer decrements (assuming intrinsic ratios suitable for AGN) suggest that these galaxies should not have high dust extinction \citep{2020ApJ...905..166L, 2021ApJ...911...70B}. However, the E(B–V) values inferred from the Pa$\alpha$/Pa$\beta$ ratio (Table \ref{tab:extinction}) for J0906 and J0954 are significantly higher; these were derived assuming an intrinsic Pa$\alpha$/Pa$\beta$ ratio of 2.133, more appropriate for H II regions. None of the four targets show strong large-scale spatial variations between spaxels in their extinction maps.

We also examine flux ratios between CLs of the same ion species as seen in Figure \ref{fig:ratios}.  While we do not observe clear systematic trends in the absolute CL fluxes across our sample, the measured line ratios are broadly consistent with those reported in ground-based observations of more massive galaxies. In particular, for commonly identified CLs such as [Si\,X]/[Si\,VI] and [S\,IX]/[S\,VIII], we find a tentative trend toward smaller line ratios in the AGN dwarfs (Figure \ref{fig:ratios}) compared to higher-mass systems. Although weak, this may indicate a subtle dependence of CL excitation conditions on black hole mass and on less extreme ionizing continua in low-mass systems. Additional ratios using other CLs can be constructed from the fluxes in Appendix \ref{appendix:fluxes}.

\subsection{Coronal line region in dwarf galaxies}\label{ref:CLR}

Determining the size and physical nature of the region where these CLs are formed (the coronal line region; CLR) has long been a challenge. Early work placed the CLR in a transitional zone between the broad-line region (BLR) and narrow-line region (NLR) \citep{1981MNRAS.195..787P, 1984MNRAS.208..347P, 1988ApJS...67..373E}, but subsequent observations showed that CLs in massive galaxies can extend to hundreds of parsecs, implying that the CLR can span a substantial fraction of the NLR \citep{2005MNRAS.364L..28P, 2006ApJ...653.1098R, 2011ApJ...739...69M, 2021ApJ...920...62N}.

The extent of a given CL depends on several coupled factors (density, radiation field, abundances, dust, etc.). A full treatment requires detailed photoionization modeling, which we will present in a forthcoming paper. Here we focus on empirical trends in the measured extents.

We use Kendall's $\tau$ (appropriate for our small sample) to test for correlations between ionization potential (I.P.) and the extents (determined using the method listed in  Section~\ref{sec:analysis}). The element-wise results are weakly consistent with smaller extents for higher-ionization lines, but are not statistically significant (likely due to the small sample size). Si shows the strongest possible trend ($\tau = -0.64$, $p = 0.055$), S is also negative ($\tau = -1.00$, $p = 0.083$), and Mg and the remaining species show negative $\tau$, while Ca shows no trend ($\tau = -0.11$, $p = 1.00$). The Ca lines despite their relatively low I.P. compared to many other CLs, have among the smallest extents. This might possibly be linked to the strong depletion of Ca onto dust grains at larger radii \citep{2004ApJS..153...75G, 2024ApJ...976..130M, 2025ApJ...984..170M}, which reduces the gas-phase Ca available for ionization. Pooling all lines in a stratified Kendall’s $\tau$ that preserves galaxy identity yields $\tau = -0.20$ ($p = 0.14$, 95\% CI [$-0.47$, $0.10$]), which is consistent with the same trend but not significant, which may be due to the fact that mixing different elements introduces abundance and depletion effects. Overall, the statistics point toward decreasing spatial extent with increasing I.P., but a larger sample is required for a definitive constraint.

Several of the stronger lines, including [Si\,VI], [Si\,VII], [Mg\,VIII], [Mg\,IV], and [Ar\,VI], show larger extents than other CLs and frequently exhibit a second, broader kinematic component. Where multiple components are present, the secondary components have  smaller spatial extents than the primary component.

If these weak trends turn out to be real, they might be interpreted in the context of photoionization theory and the density structure of gas around AGN \citep{1997ApJ...487..122F, 2011ApJ...743..100R, 2013MNRAS.430.2411M}. In ionization-bounded conditions, the CLR size is limited by the available ionizing photons: high-I.P. CLs form closer to the nucleus, while lower-I.P. CLs can be sustained at larger radii. In gas-density-bound conditions, ionizing photons are plentiful but the gas density is too low for efficient collisional excitation into the upper levels, linking the spatial distribution of a line to its critical density $n_{\mathrm{crit}}$. For $n < n_{\mathrm{crit}}$, line emission scales roughly with $n^2$, so CLs with higher $n_{\mathrm{crit}}$ are preferentially emitted in denser inner regions and lower-$n_{\mathrm{crit}}$ lines can extend farther out.

We find no significant correlation between CL extent and $n_{\mathrm{crit}}$. This suggests that the CL extents might be primarily governed by the ionizing radiation field rather than by gas-density limitations. Combined with the results of Section~\ref{sec:Source of ionizing mechanism}, this supports a picture in which AGN photoionization dominates the production and radial extent of CLs.

\begin{figure}[h!]
     \centering
     \includegraphics[width=1\linewidth]{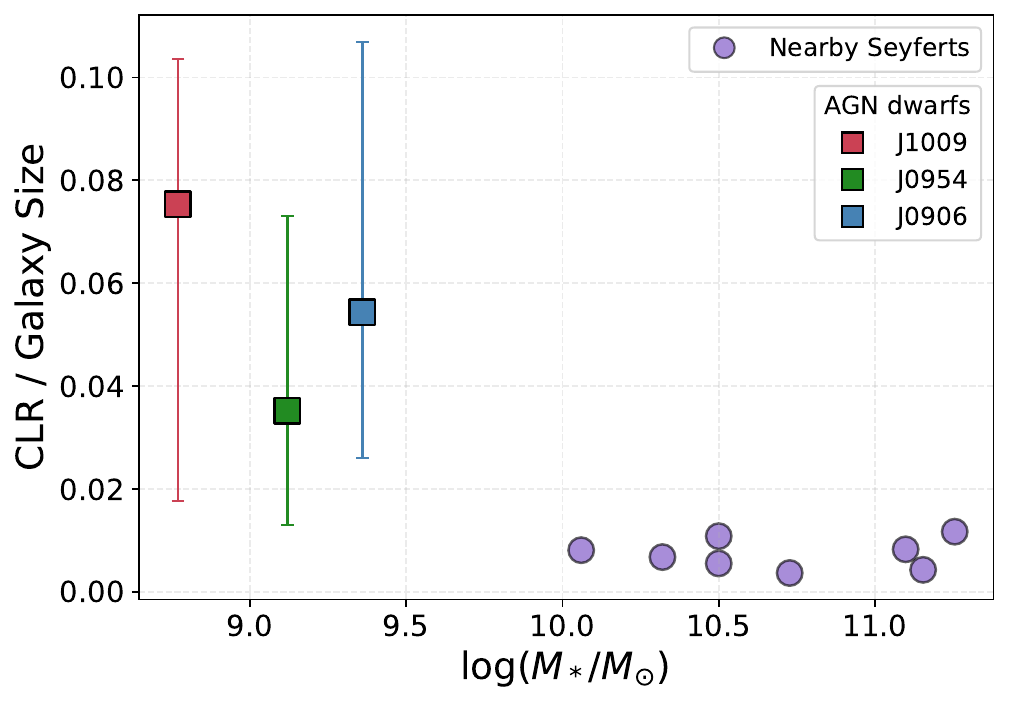}
     \caption{Ratio of the average NIR CL extents to the SDSS $r$-band isophotal radius as a function of stellar mass, comparing our dwarf galaxies (colored squares) to more massive systems (purple circles; \citealt{2011ApJ...739...69M, 2017MNRAS.470.2845R, 2018MNRAS.481L.105M}). In dwarf galaxies, the CLR can span up to $\sim10\%$ of the galaxy size, compared to $\sim1\%$ in massive galaxies.}
     \label{fig:CLR}
\end{figure}

\begin{figure*}[ht!]
    \centering
    \includegraphics[width=1\linewidth]{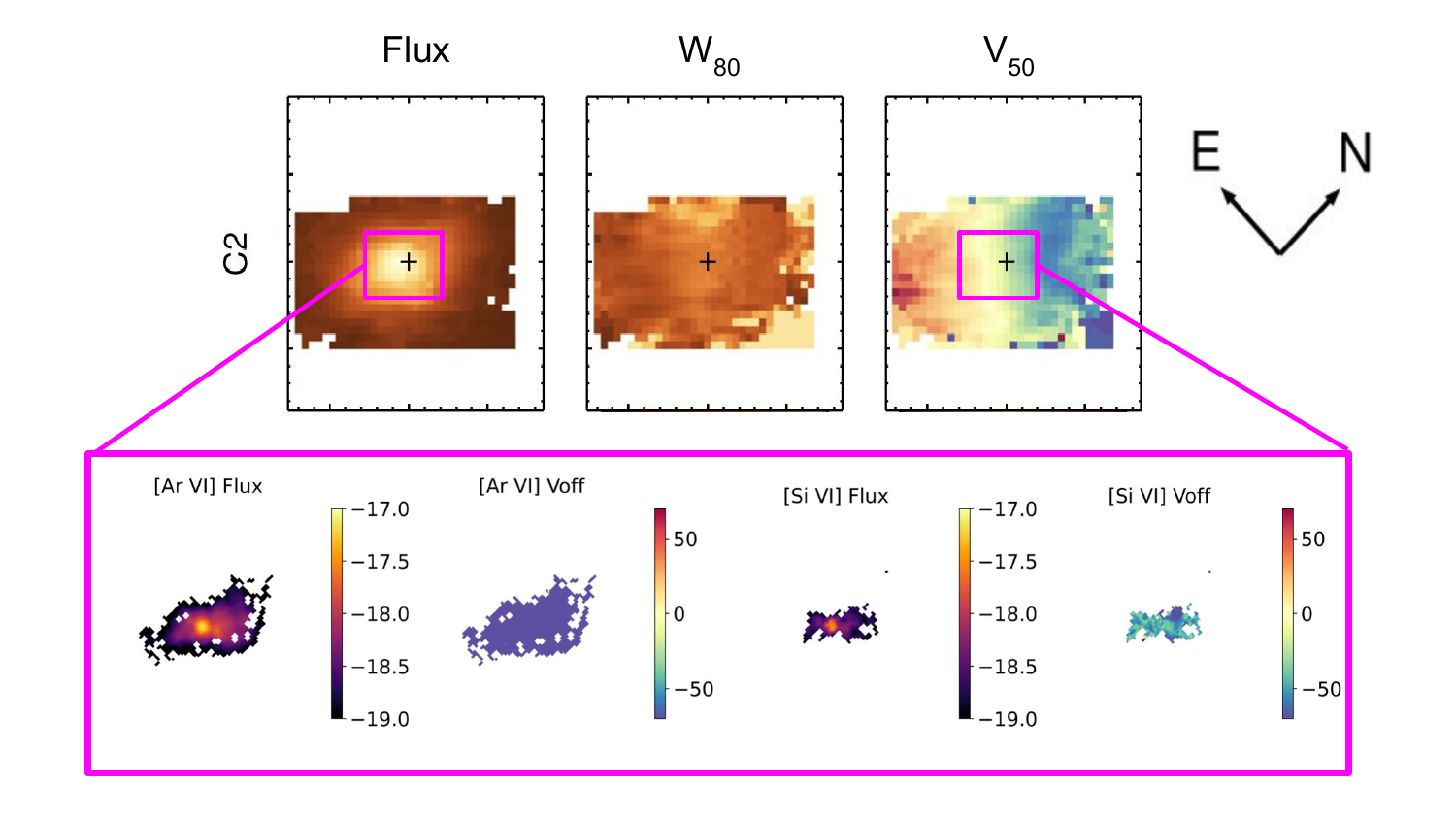}
    \caption{[O\,III] ionized gas outflow extent in J1009 from KCWI \citep{2020ApJ...905..166L} (top panel) and the corresponding CL extents for two of the CLs from JWST/NIRSpec. The flux maps are in log units of erg s$^{-1}$ cm$^{-2}$ \AA$^{-1}$ and the velocity maps are in km s$^{-1}$, with the same scale for the CLs as the [O\,III]. We find that the CL flux maps across all the targets are extended in the same direction as the [O\,III] ionized gas.}
    \label{fig:oiii_cl_extent}
\end{figure*}

The average CL extent in our sample is $\lesssim 0.5$ kpc (Figure~\ref{fig:Extents}). For J0954, the outflow radius of $\sim 0.5$ kpc inferred from blue-shifted UV absorption (C\,II, C\,IV, Si\,II, Si\,IV) by \citealt{2024ApJ...965..152L} is consistent with our coronal-line extents, indicating that the CLs might trace the ionized outflow. When multiple components are present, the broader component has typical extents of $\sim0.1$ kpc or less. We can compare our CL extents with measurements in massive galaxies, assuming those have similar sensitivity to our measurements \citep{2011ApJ...739...69M, 2005MNRAS.364L..28P, 2006ApJ...653.1098R}. The absolute CL extents in our dwarfs are similar, but because the hosts are smaller, the CLR appears to occupy a larger fraction of the galaxy ($\sim$10\% versus $\sim1\%$). However, optical CLs in massive galaxies have been shown to be more extended (0.7--3 kpc; \citealt{2017MNRAS.470.2845R, 2021ApJ...920...62N, 2025MNRAS.538.2800R}) so our NIR-based comparison is conservative.

\begin{figure*}
    \centering
    \includegraphics[width=1\linewidth]{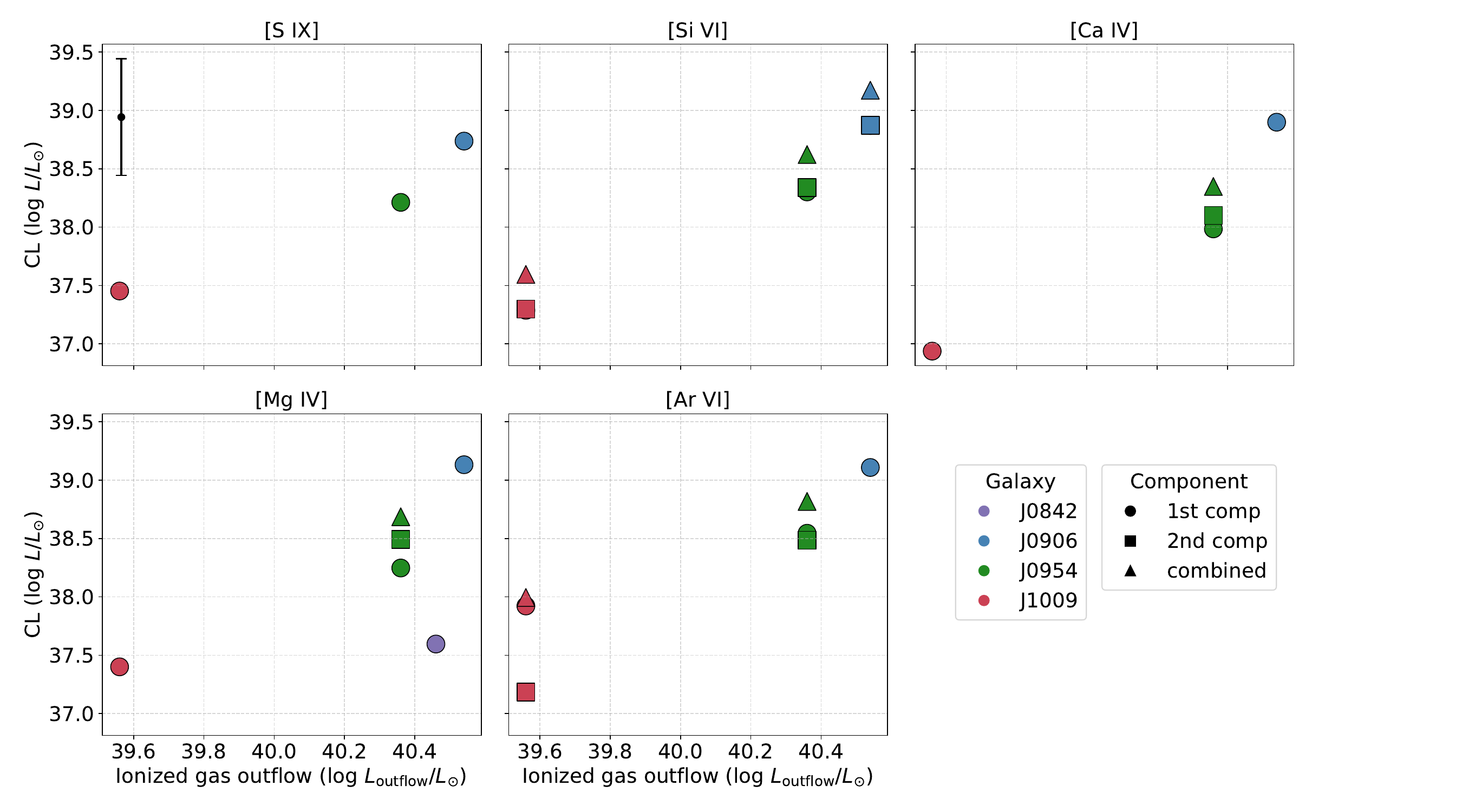}
    \caption{Comparison of the outflow luminosity (calculated from \citealt{2020ApJ...905..166L}) with CL luminosities. The error is indicated in black in the first panel. Across CL species, the galaxy with the lowest [O\,III] outflow luminosity (J1009) also exhibits the lowest CL luminosities. }
    \label{fig:energetics_outflow}
\end{figure*}

\begin{figure*}
    \centering
    \includegraphics[width=1\linewidth]{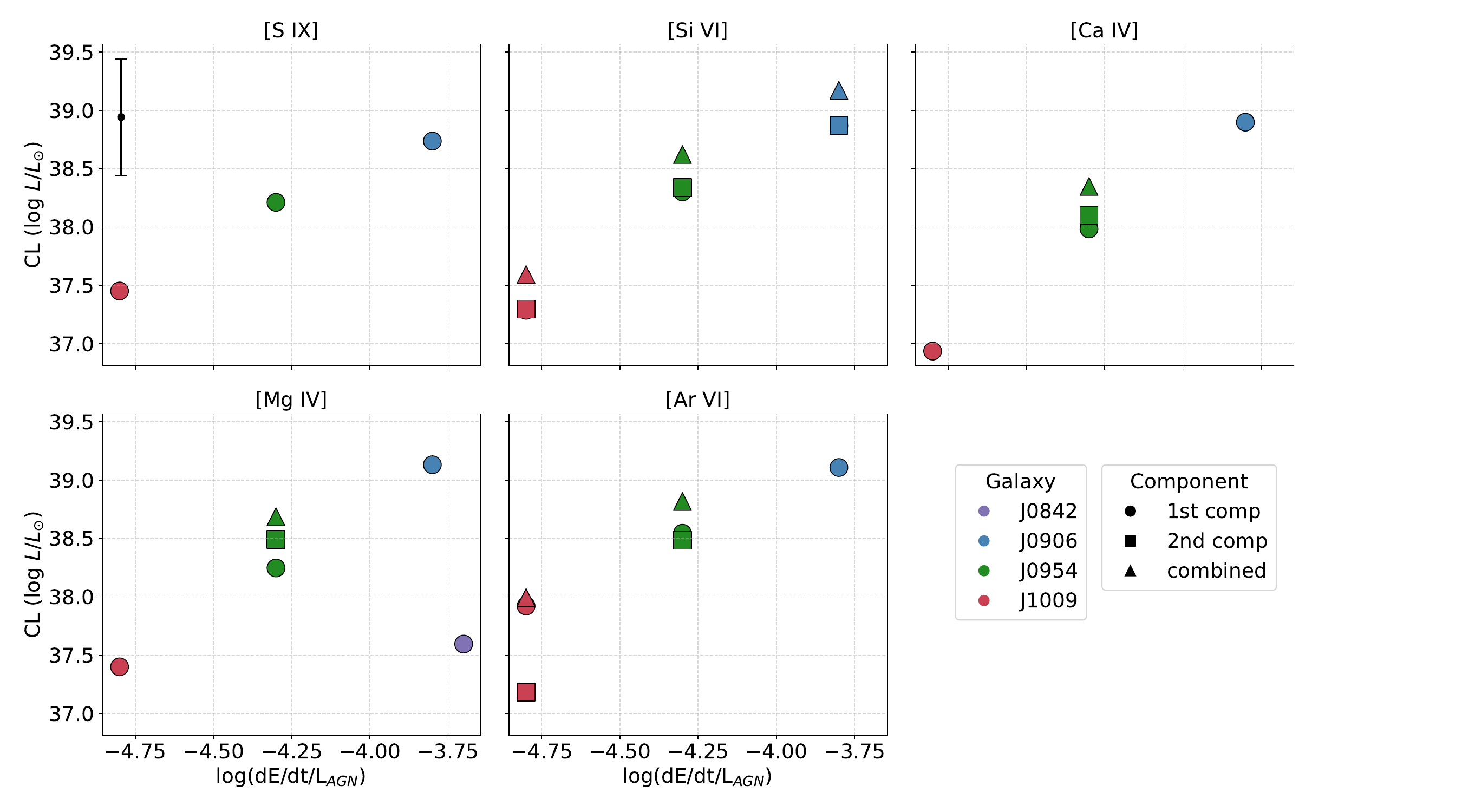}
    \caption{[O\,III] outflow kinetic energy (normalized by L$_{\mathrm{AGN}}$) vs. CL luminosity. Typical errors are indicated in black in the first panel. We find that the most energetic [O\,III] outflows tend to have the most luminous CL emission across all CLs.}
    \label{fig:lum_dedt}
\end{figure*}

\subsection{Correlation with ionized-gas outflows} \label{ionized}

Although the AGN accretion disk is expected to generate abundant photons capable of producing CLs, recent systematic surveys investigating optical CLs in large samples of nearby galaxies have found such emission to be exceedingly rare within the general galaxy population. This rarity persists even among galaxies identified as AGNs using standard narrow-line ratio diagnostics, at current survey sensitivities \citep{2021ApJ...922..155M, 2021ApJ...920...62N, 2022ApJ...936..140R, 2023ApJ...945..127N}. Furthermore, these surveys indicate that CL emission preferentially appears in galaxies with low dust extinction \citep{2004ApJS..153...75G, 2025ApJ...984..170M}, suggesting that dust may contribute to suppressing CL emission.

Recent photoionization models by \citealt{2024ApJ...976..130M} show that gas-phase depletion of refractory elements onto dust grains, along with dust absorption of ionizing photons, can significantly impact CL emission, reducing line luminosities by up to three orders of magnitude compared to dust-free conditions. Consequently, strong CL emission likely occurs only in environments where sufficient grain destruction takes place within highly ionized gas \citep{2003AJ....125.1729N}. All dwarf galaxies with CL in this sample also exhibit low levels of dust extinction in the central regions, at least based on their Balmer decrements (Table \ref{tab:extinction}). 

All galaxies in this work show some of the fastest ionized-gas outflows reported in dwarf galaxies, as traced by their broad, blue-shifted [O\,III] emission-line profiles \citep{2019ApJ...884...54M, 2020ApJ...905..166L}. The [O\,III] outflow component contributes 5–30\% of the total [O\,III] flux in these galaxies \citep{2019ApJ...884...54M}. Both J0906 and J0954 require three Gaussian components to model the [O\,III] emission lines, revealing a broad component with an average FWHM of $\sim$800 km s$^{-1}$. J0906 and J0954 have maximum outflow velocities of $-$150 and $-$80 km s$^{-1}$, respectively, where the negative sign indicates blue-shifted velocity centroids relative to the systemic velocity. J1009 has a maximum outflow velocity of $-$60 km s$^{-1}$ (see also Section \ref{sec:OIII outflows} and Table \ref{tab:tab1} for more details on the ionized gas outflow properties). The warm ionized-gas mass outflow rates range from $\sim 1$–$3 \times 10^{-2}$ M$_\odot$ yr$^{-1}$, and the kinetic energy outflow rates vary between $1.5 \times 10^{38}$ to $8 \times 10^{39}$ erg s$^{-1}$, which are comparable to those of more luminous AGN in massive systems when normalized by AGN bolometric luminosities.

Outflows could enhance CL emission by clearing out dust from the interior of the galaxies. This was previously suggested in a study of bulgeless galaxies with AGN, where \citealt{2022ApJ...931...69B} found that all the galaxies that have AGN-driven outflows also show CL emission, suggesting a  correlation between the two. In a systematic study of galaxies with ionized-gas outflows, \citealt{2025ApJ...984..170M} showed that galaxies with CL emission exhibit a significantly higher incidence and luminosity of ionized outflows, traced by [O\,III] $\lambda$5007, compared to a control sample of non-CL-emitting galaxies. They also found that there exists strong correlations between CL properties (luminosity and FWHM) and outflow velocity. They find that the CL emitters have systematically lower intrinsic extinction toward the ionized gas compared to the controls, supporting a scenario in which dust destruction in AGN-driven outflows facilitates efficient CL emission. Although there is a possibility that this correlation is a by-product of AGN power, \citealt{2025ApJ...984..170M} do not find a correlation between the total AGN bolometric luminosity and the CL luminosity (which we also find in this work.)

In our analysis, we also find several correlations between infrared CL properties and the characteristics of ionized gas outflows traced by [O\,III] \citep{2020ApJ...905..166L} in our sample galaxies. The detected CLs are spatially extended along the same direction as the outflows detected in [O\,III] (Figure \ref{fig:oiii_cl_extent}). The extended CLs  also exhibit a largely biconical morphology.

\begin{figure}[h!]
    \centering
    \includegraphics[width=1\linewidth]{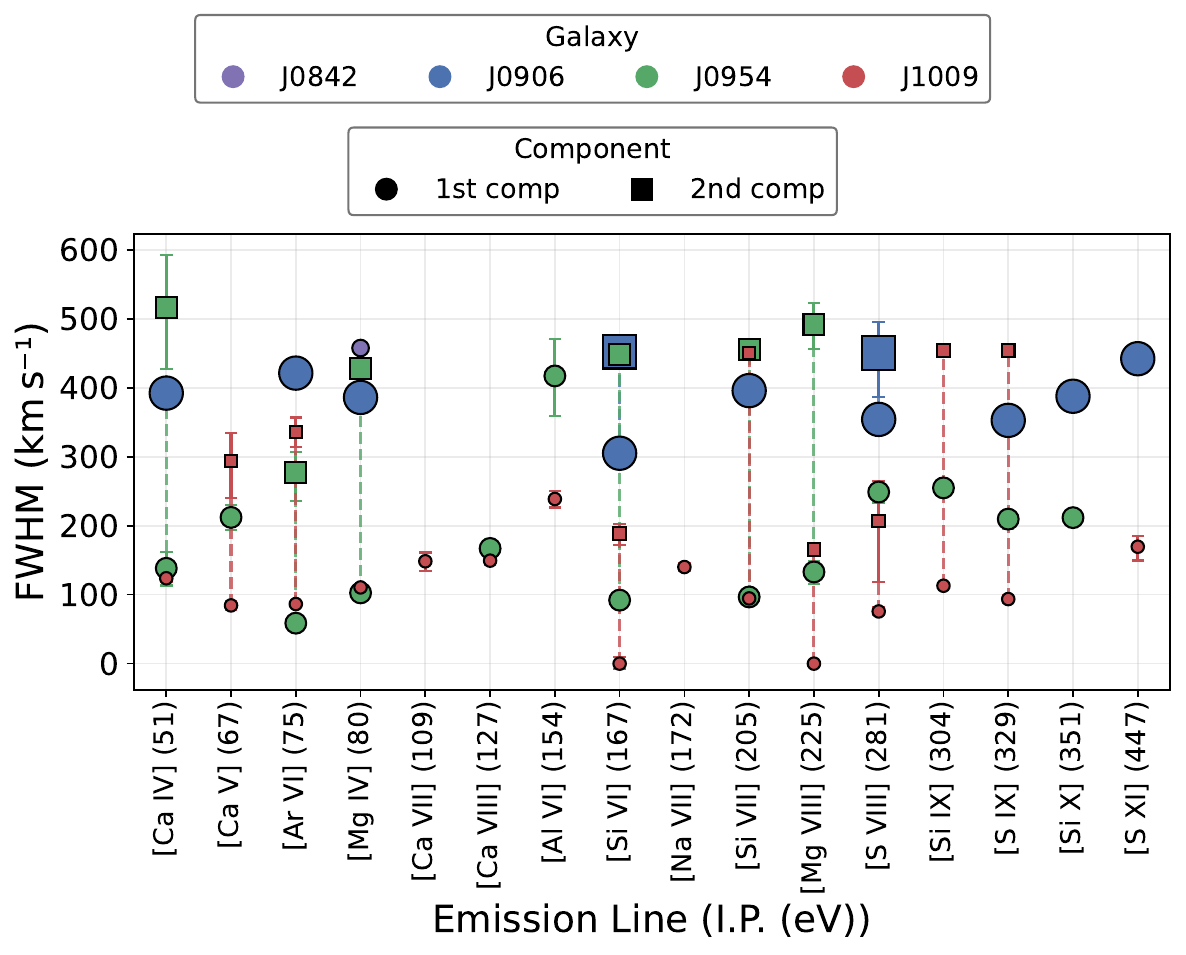}
    \caption{FWHM of the observed CL with the I.P. The size of the marker corresponds to the speed of the [O\,III] ionized gas outflows in the corresponding galaxy. There is no correlation between the I.P. and FWHM in our sample. }
    \label{fig:FWHM_I.P.}
\end{figure}

Furthermore, we apply a Kendall's $\tau$ correlation test to determine if a significant correlation exists between the various quantities, such as the CL luminosity and CL FWHM and ionized gas outflow luminosity, [O\,III] FWHM and the [O\,III] outflow energetics. The ionized gas outflow luminosity is calculated from the outflow masses given in Table 4 of \citealt{2020ApJ...905..166L} using Equation 7 in the same work. The CL luminosities and FWHM are calculated from the integrated spectra using a 0.3\arcsec central aperture for all the detected CL (we show a subsample of the different CL in Figures \ref{fig:energetics_outflow} and \ref{fig:lum_dedt}, but the correlations are present across all the CLs). We find the strongest correlations (median Kendall's $\tau$ coefficient = 0.42 and p-value = 0.05) between the CL luminosity and the AGN outflow luminosity (Figure \ref{fig:energetics_outflow}) as well as between the CL luminosity and the most energetic outflows, given as the kinetic energy injection rate (dE/dt) normalized by the AGN bolometric luminosity (Figure \ref{fig:lum_dedt}). We do not find a significant correlation between the total AGN bolometric and [O\,III] luminosities and the CL luminosities. We also find a significant correlation (median Kendall's $\tau$ coefficient $>$ 0.4 and p-value $<$ 0.05) between CL luminosity and the FWHM of the [O\,III] outflow component. We do not find a significant correlation between the FWHM of CL and that of the [O\,III] outflow component, which agrees with the results from \citealt{2025ApJ...984..170M}.

While some previous studies have suggested that CLs with higher I.P. tend to exhibit broader profiles \citep{2002ApJ...579..214R, 2023ApJ...942L..37A}, we do not find a statistically significant correlation between either FWHM or W$_{80}$ and the I.P. or the critical density (n$_{\mathrm{crit}}$) of the lines (Figure \ref{fig:FWHM_I.P.}). However, we do find that targets with larger W$_{80}$ values in the broad [O\,III] component also show correspondingly large W$_{80}$ values across their CLs (Figure \ref{fig:FWHM_I.P.}).

Although the limited size of our sample makes it difficult to characterize broader trends, our results are consistent with previous studies suggesting that outflows may play a role in dust destruction, thereby enabling efficient CL detection.

\subsection{Outflows traced by Coronal Lines}\label{sec:Hot ionized gas outflows}

We find that several coronal lines (CLs) in the observed dwarf galaxies require a second, broad component. This may indicate that the coronal-line region (CLR) arises closer to the central AGN than the narrow-line region (NLR), leading to intrinsically broader profiles \citep{2002ApJ...579..214R, 2010MNRAS.405.1315M}. Broad CLs have also been reported in tidal disruption events (TDEs; \citealt{2011ApJ...740...85W}), but the lack of fading between the Keck/NIRES observations taken in 2019 and JWST/NIRSpec observations taken in 2024 (see Section \ref{sec:coronal line detections}) makes a TDE origin unlikely in our case. In massive AGN, CLs are often associated with outflows, and their typically blue-shifted, broadened profiles are frequently interpreted as signatures of outflowing gas \citep{2011ApJ...739...69M, 2013MNRAS.430.2411M, 2017MNRAS.470.2845R, 2018MNRAS.481L.105M, 2021MNRAS.506.3831F,  2022A&A...665A..55S}.

\citet{2021ApJ...911...70B} detected a secondary broad component in the [Si\,VI] emission lines of J0906, J0954, and J1009 using Keck/NIRES. They measured velocity offsets up to $-$600 km s$^{-1}$ and maximum W$_{80}$ values of $\sim$1000 km s$^{-1}$ for the broad component, strongly suggestive of outflowing coronal gas. We do not recover comparably high velocities in [Si\,VI] with JWST/NIRSpec. However, individual spaxels in the NIRSpec IFU cubes are affected by “wiggles” (Section \ref{sec:Observations and Data reduction}), which complicates separating weak, broad wings from spectral artifacts and likely leads us to underestimate the outflow velocities. Nevertheless, the presence of a broad component in both the CLs and the [O\,III] lines in the ground-based data motivates a closer examination of whether even the highest-ionization gas participates in the outflow.

The broad CL components in this work show velocity offsets of $-$150 to 150 km s$^{-1}$ and W$_{80}$ values of $\sim$300–500 km s$^{-1}$. These values should be treated as lower limits, since the outflow wings are difficult to disentangle from the “wiggles.” Overall, the broad CL components have smaller velocities than lines such as [S\,III] $\lambda$0.9533 $\mu$m and He\,II $\lambda$1.0125 $\mu$m, which have similar ionization potentials and electron densities to [O\,III]. A more detailed outflow analysis using additional emission lines will be presented in a future study.

\begin{figure*}[ht!]
    \centering
    \includegraphics[width=0.9\linewidth]{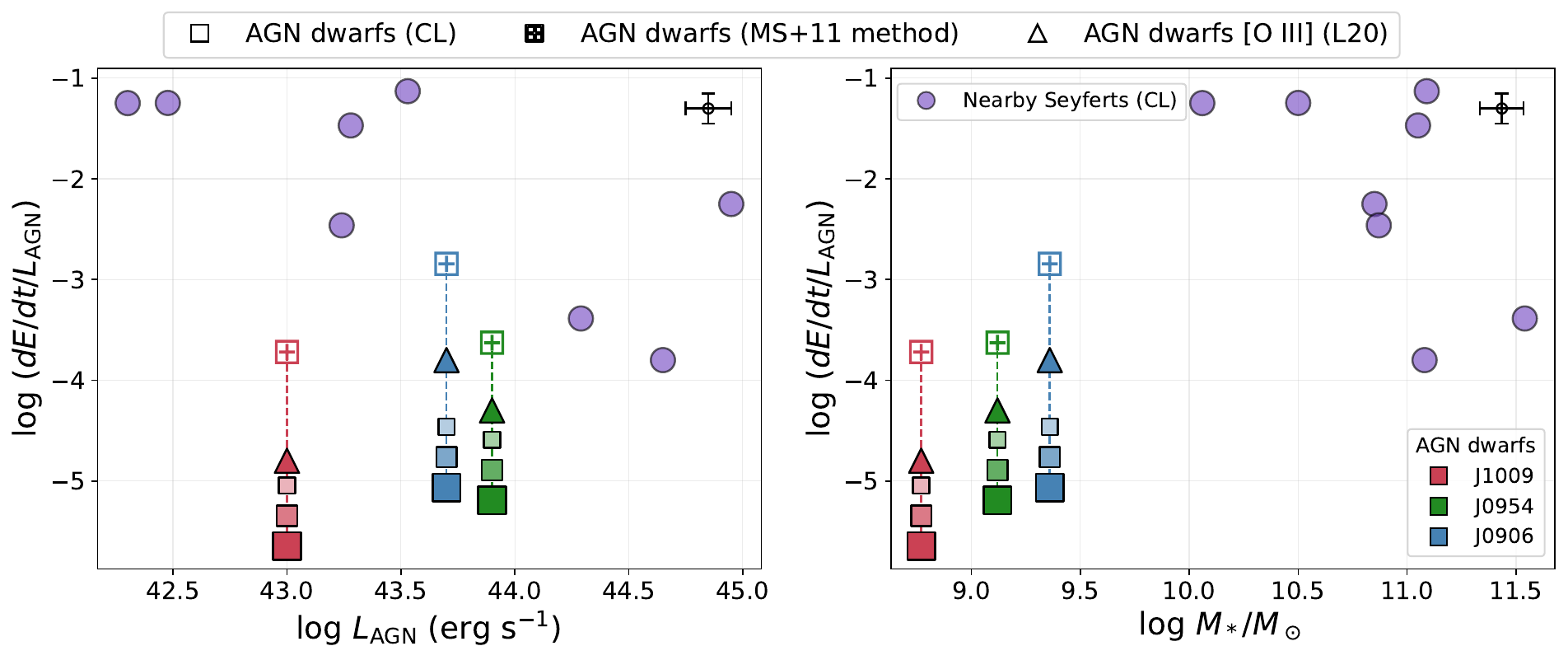}
    \caption{Kinetic energy outflow rates (see Equation (5)) normalized to the AGN luminosity as a function of AGN luminosity (left) and stellar mass (right). The colored (red, green, blue) points indicate AGN dwarfs and the purple circles indicate measurements from nearby Seyferts \citep{2011ApJ...739...69M, 2017MNRAS.470.2845R, 2018MNRAS.481L.105M} with spatially resolved NIR CL measurements. The squares indicate measurements for [Si\,VI] using Equations (2) - (5), with increasing sizes from big to small (upwards in the kinetic energy outflow rate) are the values assuming Z$_{\odot}$, 0.5\,Z$_{\odot}$ and 0.25\,Z$_{\odot}$ respectively.} The triangles are measurements from [O\,III] \citep{2020ApJ...905..166L}. The squares with crosses indicate energetics calculated assuming a filling factor using equations from \citealt{2011ApJ...739...69M}, also used in \citealt{2017MNRAS.470.2845R}. Across the sample, coronal-line–based outflow energetics estimated using the filling factor (squares with crosses) approximation are 1.0–2.0 dex higher than those estimated using the [Si\,VI] luminosities (filled squares).
    \label{fig:energetics}
\end{figure*}

Assuming the secondary component seen in the CLs is an outflow, we can derive standard estimates of energetics, including the mass outflow and kinetic energy  rates. Previous studies on CL energetics \citep{2011ApJ...739...69M, 2017MNRAS.470.2845R} account for the velocities and the extents of the outflow. However, they do not measure the mass from the luminosities of the CL, but instead assume a certain filling factor and make use of the hydrogen mass alone. The unknown filling factor leads to large uncertainties in the calculated energetics. In this work, we make use of [Si\,VI] luminosities in order to estimate the amount of gas mass present in the outflow, and use that to determine the outflow energetics (see Appendix \ref{appendix:outflow} for a more detailed derivation). We use the following equation to derive the outflow mass for a steady state-outflow:

 \begin{equation}
     {M}_{\rm{coronal\hspace{0.1cm}gas}} = \frac{160}{(Z/Z_\odot)}\hspace{0.1cm}M_{\odot} \Big(\frac{\hspace{0.1cm}L_{\mathrm{[Si\,VI],\hspace{0.1cm} ext.corr}}}{10^{35}\hspace{0.1cm}erg\hspace{0.1cm} s^{-1}}\Big)\Big(\frac{n_e}{100\hspace{0.1cm}cm^{-3}}\Big)^{-1}
 \end{equation}
where $L_{\mathrm{[Si\,VI],\hspace{0.1cm} ext.corr}}$ is the extinction-corrected [Si\,VI] luminosity and $n_e$ is the electron density, measured from the [S\,II] $\lambda\lambda$6716, 6731 line ratios, obtained from Gemini/GMOS spectra or Keck/LRIS spectra \citep{2020ApJ...905..166L}. We caution that the $n_e$ we adopt from the [S\,II] doublet likely traces lower-ionization NLR gas rather than the coronal line–emitting region itself \citep{2019MNRAS.486.4290B, 2020MNRAS.498.4150D}. If the CLR is denser than the [S\,II]-emitting gas, our assumed $n_e$ is underestimated and the inferred outflow masses and rates should be treated as upper limits. Additionally, the value of the outflow mass will increase proportionally if we assume sub-solar metallicities for $\mathrm{(Z/Z_\odot)}$.

\begin{deluxetable*}{cccccccccc}
\tablecaption{Energetics of the gas, traced by the [Si\,VI] line}
\tablehead{\colhead{Name} & \colhead{Comp} & \colhead{n$_e$}& \colhead{R$_{out}$} & \colhead{Avg. v$_{\mathrm{off}}$} & \colhead{Avg. W$_{80}$} & \colhead{log(M$_{out}$)} & \colhead{log[(dM/dt)]} & \colhead{log[(dE/dt)]} & \colhead{log[(c dp/dt)]}\\
& & \colhead{(cm$^{-3}$)} & \colhead{(kpc)} & \colhead{(km s$^{-1}$)} & (km s$^{-1}$) & (M$_{\odot}$) &(M$_{\odot}$ yr$^{-1}$) & (erg s$^{-1}$) & (L$_{\odot}$)\\
\colhead{(1)} & \colhead{(2)} & \colhead{(3)}& \colhead{(4)} & \colhead{(5)} & \colhead{(6)} & \colhead{(7)} & \colhead{(8)} & \colhead{(9)} & \colhead{(10)}}
\startdata
             J0906  & 1 &  570 $\pm$ 360 & 0.203 & 30 $\pm$ 16  &  337 $\pm$ 43 & 5.27$^{+0.6}_{-0.9}$ & $-$1.70$^{+0.5}_{-0.4}$ & 38.63$^{+0.5}_{-0.5}$ & 7.67$^{+0.3}_{-0.5}$ \\
             J0954  & 1 &  470 $\pm$ 80 & 0.254 & $-$6 $\pm$ 12 & 138 $\pm$ 39 & 4.69$^{+0.1}_{-0.2}$ & $-$2.73$^{+0.1}_{-0.2}$ &	36.76$^{+0.1}_{-0.2}$ &	6.23$^{+0.1}_{-0.2}$ \\
                    & 2 &  470 $\pm$ 80 & 0.099 & 60 $\pm$ 24& 405 $\pm$ 43 & 4.34$^{+0.3}_{-0.3}$ & $-$1.81$^{+0.3}_{-0.3}$ &	38.70$^{+0.3}_{-0.3}$ &	7.99$^{+0.3}_{-0.3}$ \\
             J1009  & 1 &  150 $\pm$ 60 & 0.254 & $-$45 $\pm$ 13 & 145 $\pm$ 40&	4.50$^{+0.2}_{-0.2}$ & $-$2.75$^{+0.2}_{-0.2}$ &	37.07$^{+0.2}_{-0.2}$ &	6.59$^{+0.2}_{-0.2}$ \\
                    & 2 &  150 $\pm$ 60 & 0.037 & $-$45 $\pm$ 13 &343 $\pm$ 35& 3.65$^{+0.2}_{-0.2}$ &	$-$3.01$^{+0.2}_{-0.2}$ &	37.35$^{+0.2}_{-0.2}$ &	6.13$^{+0.2}_{-0.2}$\\
\enddata
\tablecomments{\footnotesize  Column (1): Short name of the target. Column (2): Individual outflow components from the best fits. Column (3) Electron density measured from the [S\,II] $\lambda\lambda$6716, 6731 line ratio based on the total line flux from the spatially integrated, Gemini/GMOS spectra or Keck/LRIS spectra (see \citealt{2020ApJ...905..166L}). Column (4): Outflow radius adopted in the calculation of mass, momentum, and kinetic energy outflow rates when the outflows are spatially resolved. Column (5): Average offset velocity calculated from the best fits. Column (6): Average W$_{80}$ calculated from the best fits. Column (7): Ionized gas mass of the corresponding outflow component.  Column (8): Ionized gas mass outflow rate of the corresponding outflow component. Column (9): Ionized gas kinetic energy outflow rate of the corresponding outflow component. Column (10): Ionized gas momentum outflow rate of the corresponding velocity component}
\end{deluxetable*}\label{table:energetics}
The calculations of the mass, momentum, and kinetic energy outflow rates depend on the spatial extent of the outflows which we calculate in Section \ref{sec:Determining the extent of the CL} and discuss in Section \ref{ref:CLR}. Specifically, the energetics are calculated by summing up quantities over individual spaxels:

\begin{equation}
    \mathrm{dM/dt} = \sum dm/dt = \sum \frac{m_{out} v_{\mathrm{off,out}}sec\theta}{R_{out}}
\end{equation}
\begin{equation}
    \mathrm{dp/dt} = \sum(v_{\mathrm{off,out}}sec\theta)dm/dt
\end{equation}
\begin{equation}
    \mathrm{dE/dt} = \frac{1}{2} \sum[(v_{\mathrm{off,out}}sec\theta)^2 + 3 \sigma_{out}^2]dm/dt,
\end{equation}
where $\mathrm{m_{out}}$ is the outflowing mass in each spaxel, $\mathrm{v_{off,out}}$ is the value of v$_{\mathrm{off}}$ in each spaxel, and $\sigma_{out}$ is the velocity dispersion calculated as W$_{80}$/2.73. $\mathrm{R_{out}}$ is the radial extent of the outflow. $\theta$ is the angle between the velocity vector of the outflow in 3D space and the line of sight, defined as $\theta$ = sin$^{-1}$(r$_{\mathrm{spaxel}}$/$\mathrm{R_{out}}$).

The calculated energetics, along with the average velocity offsets, W$_{80}$ and the extents R$_{out}$ used in the calculations of the energetics are given in Table \ref{table:energetics}.

In Figure \ref{fig:energetics}, we plot the mass and kinetic energy outflow rates estimated for the outflowing gas traced by the CLs as a function of mass and AGN luminosity. We also compare the outflow rates from CL to the ionized [O\,III] gas measurements and find that the CL outflow rates are 1 dex lower, which could be due to underestimating the W$_{80}$ values due to the de-wiggling process (see Section \ref{sec:wiggles}). We also compare the energetics measured in this work to energetics calculated using [Si\,VI] outflows from ground-based low-z Seyferts \citep{2011ApJ...739...69M, 2017MNRAS.470.2845R, 2018MNRAS.481L.105M}. Since the mass outflow rates in these works were calculated assuming a filling factor, we also calculated the energetics from the AGN dwarfs using the equations presented in \citealt{2011ApJ...739...69M} (indicated by the boxes with a cross). We assume a filling factor (0.001) similar to \citealt{2011ApJ...739...69M} and find that this measurement over estimates the values for energetics by at least two orders of magnitude. Even if we assume higher n$_e$ and filling factor values suitable for dwarf galaxies, the energetics are still found to be higher. This could indicate that the filling factors for CL are much lower than previously assumed.

Because the highly ionized phase of the gas traced by CLs appears neither  fast nor radially extended, it likely cannot expel gas from the galaxy. However, its energetics may be sufficient to disturb gas in the central regions, potentially contributing to feedback-driven gas redistribution that has been invoked to produce cored (rather than cuspy) dark-matter profiles in dwarf galaxies \citep{2010AdAst2010E...5D}. They might also accelerate gas and redistribute mass in the central, high–gravitational-potential regions, potentially driving more extended, lower-ionization outflows (e.g., [O\,III]).

\section{Summary and Conclusions}\label{sec:summary}

Using JWST/NIRSpec, we undertake the \textit{first} spatially resolved examination of NIR coronal lines in dwarf galaxies with AGN and ionized-gas outflows to address the challenges in unambiguously identifying AGN activity and assessing AGN-driven feedback in these low-mass systems. We generate spatial maps of the flux and velocity of the different coronal lines, allowing us to examine in detail the influence of AGN feedback, as traced by the coronal lines, on the host dwarf galaxies. Our selected targets host [O\,III]-traced ionized outflows and have prior evidence for CL emission, and the present study therefore focuses on a currently limited subset of dwarf galaxies. However, growing evidence for outflows in AGN dwarf galaxies \citep{2025ApJ...979...26S} suggests that broader, more representative extensions of this work will be feasible in the future. We present our main results as follows:
\begin{itemize}
\item 16 different species of near-infrared coronal lines were detected across four dwarf galaxies, tripling the number of detected CL transitions in dwarf galaxies compared to previous ground-based studies. The galaxy with the lowest stellar mass (J1009, M$_*$ = 10$^{8.77}$ M$_{\odot}$, about 2.5 times less massive than the other dwarf galaxies in the observed sample) exhibits the maximum number of CLs (15) among all the galaxies, indicating that black holes in small dwarf galaxies are capable of producing the ionization required to form CLs.
\item Based on shock and AGN models, we find that AGN photoionization is largely responsible for producing CLs, although shock excitation may contribute to the large extents of some CLs, especially if we assume lower metallicities. J1009, in particular, is dominated by AGN photoionization with minimal to no contribution from shock ionization to producing the CLs. The models also suggest that J1009 is hosting a highly accreting, low-mass black hole that is capable of producing numerous CL with I.P.s as high as 447 eV.
\item The average extents of the CLs across the four dwarf galaxies are $\sim$0.5 kpc, which is similar to the extents found tracing NIR CL in massive galaxies. The small sizes of the dwarf galaxies in this sample mean their CLRs can occupy $\sim$ 10\% of the host galaxy, about ten times the fractional extent seen in massive galaxies. In J0954, an outflow radius of $\sim$0.5 kpc estimated from blue-shifted UV absorption \citep{2024ApJ...965..152L} agrees with the CL extents, supporting CLs as tracers of the ionized outflow.
\item The presence, extent, and luminosity of the CLs correlate with the properties of the ionized gas outflows traced by [O\,III]. In galaxies with spatially resolved CL emission, CLs exhibit biconical structure, align with the outflow axis, consistent with an AGN-powered outflow. The CL luminosities correlate with both the ionized gas outflow luminosities and the FWHM of the outflows. Two of the galaxies with CL detections also  have small dust extinction values, suggesting that AGN-driven outflows in these galaxies may enhance CL emission by clearing dust from the nuclear regions.
\item Several CLs exhibit wings, suggestive of a possible secondary kinematic component. If this is an outflow, energy estimates imply this highly ionized gas in these galaxies can have a dynamical impact on the gas near galaxy centers.
\end{itemize}
The results confirm that the outflows detected in the dwarf galaxies in this sample are predominantly AGN-powered and that CLs can trace their presence and impact. By revealing signatures of dust clearing, gas heating, and kinematically distinct ionized outflows, CLs may provide a direct view of how AGN-driven outflows in dwarf galaxies can exert feedback, potentially influencing their evolution. Our findings highlight the potential for CLs as diagnostics of AGN activity in low-mass, dust-poor environments, which are conditions reminiscent of galaxies in the early universe. With increased observations of AGN in faint galaxies at high redshifts \citep{2023MNRAS.525.2087B, 2023ApJ...952..142F, 2023ApJ...959...39H, 2023ApJ...954L...4K, 2023Natur.616..266L, 2024Natur.627...59M}, including detections of CLs such as [Ne\,V] in dwarf galaxies at the epoch of reionization \citep{2024MNRAS.534.2633C} using JWST, understanding the formation and detection of CLs in dwarf galaxies is critical. Given that local dwarf galaxies are crucial analogs for understanding the formation and evolution of the earliest galaxies in the high-redshift universe, where they are increasingly recognised as key drivers of reionization \citep{2024Natur.626..975A} and where AGN feedback in dwarf galaxies is gaining importance in galaxy quenching \citep{2024A&A...690A.286A}, these results can provide invaluable empirical constraints for cosmological simulations and future JWST observations aimed at unveiling the role of AGN in early galaxy evolution.

\begin{acknowledgments}
We thank the anonymous referee for their thoughtful feedback that greatly helped to improve this paper. This work is based on observations made with the NASA/ESA/CSA James Webb Space Telescope. The data were obtained from the Mikulski Archive for Space Telescopes at the Space Telescope Science Institute, which is operated by the Association of Universities for Research in Astronomy, Inc., under NASA contract NAS 5-03127 for JWST. These observations are associated with program \#3663. All the {\it JWST} data used in this paper can be found in MAST doi:\dataset[10.17909/19nr-xp61]{http://dx.doi.org/10.17909/19nr-xp61}. Support for program \#3663 was provided by NASA through a grant from the Space Telescope Science Institute, which is operated by the Association of Universities for Research in Astronomy, Inc., under NASA contract NAS 5-03127. VU acknowledges partial funding support from NSF AAG grant \#2536603 and NASA ADSPS grant \#80NSSC25K0169. MB acknowledges support from the Juan de La Cierva scholarship with reference JDC2023$-$052684$-$I, funded by MICIU/AEI/10.13039/501100011033 and from the Agencia Estatal de Investigaci\'on of the Ministerio de Ciencia, Innovaci\'on y Universidades (MCIU/AEI) under the grant ``Tracking active galactic nuclei feedback from parsec to kiloparsec scales'', with reference PID2022$-$141105NB$-$I00 and the European Regional Development Fund (ERDF). We thank Danielle Berg and Zorayda Martinez for helpful discussions regarding this work.
\end{acknowledgments}




%
\facilities{JWST(NIRSpec)}

\software{astropy \citep{2013A&A...558A..33A,2018AJ....156..123A,2022ApJ...935..167A},  JWST Calibration Pipeline \citep{2023zndo...7577320B},
          BADASS \citep{2021MNRAS.500.2871S},
          pPXF, \citep{2004PASP..116..138C},
         PyNeb \citep{2015A&A...573A..42L}, }


\appendix

\section{Outflow energetics} \label{appendix:outflow}
Spatially resolved IFU data has proven to be integral to deriving various energetics of outflows (the amount of mass carried by the outflows, the rates of mass, momentum and kinetic energy of the outflow), by providing constraints on the spatial extent of the outflow, which is a crucial factor in the calculation of the energetics. The energetics are dependent on several assumptions, such as the shape of the outflow (spherical vs. biconical). The ionized gas mass of the outflows can be calculated based on either the [O\,III] $\lambda, \lambda$4959, 5007 line luminosity or the Balmer line (H$\alpha$ or H$\beta$) luminosity of the outflowing, line emitting gas. Calculating the coronal gas mass using CLs is challenging. Previous works \citep{2011ApJ...739...69M, 2017MNRAS.470.2845R, 2018MNRAS.481L.105M, 2020ApJ...895L...9R}, typically assume a spherically symmetric outflow with a certain filling factor, with only the outflow velocities and the spatial extent as the varying parameters, and do not account for the gas mass from the luminosity of the CL. We find that this method assuming a filling factor typically overestimates the mass and kinetic energy outflow rates (Figure \ref{fig:energetics}). Since the luminosity of the CLs give us the best estimate of the coronal ionized gas mass in the galaxy, in this work, we derive the coronal ionized gas mass using the luminosity of the [Si\,VI] CL, following the methods from \citealt{2006agna.book.....O, 2020A&ARv..28....2V, 2020ApJ...905..166L}, assuming a case B recombination.

The luminosity of [Si\,VI] can be given as 
\begin{equation}
L_{\mathrm{Si\,VI}}
= \int j_{\mathrm{[Si\,VI]}}\,dV
\end{equation}

where j$_{[SiVI]}$ is the line emissivity coefficient and can be written as:

\begin{equation}
j_{\mathrm{[Si\,VI]}}
= n_e\,n_{\mathrm{[Si\,VI]}}\,\epsilon_{\mathrm{[Si\,VI]} (Te, Ne)}
\end{equation}

where n$_e$ is the electron density, $n_{\mathrm{[Si\,VI]}}$ is the number density of [Si\,VI] and $\epsilon_{\mathrm{[Si\,VI]}}$ is the line emission coefficient. 

The total coronal gas mass associated with [Si\,VI], M$_{cor}$, is primarily hydrogen and helium. Si is only a trace element. So, to convert [Si\,VI] density into a total mass, we assume a certain abundance ratio $A_{\mathrm{Si}}$:
\begin{equation}
A_{\mathrm{Si}}
= \frac{n_{\mathrm{Si}}}{n_{\mathrm{H}}}
\end{equation}
Considering the ionization fraction of [Si\,VI] $f_{\mathrm{[Si\,VI]}}$,
\begin{equation}
    f_{\mathrm{[Si\,VI]}}
= \frac{n_{\mathrm{[Si\,VI]}}}{n_{\mathrm{Si}}}
\end{equation}

We can now find the number density of hydrogen n$_{H}$:
\begin{equation}
n_{\mathrm{H}}
= \frac{n_{\mathrm{[Si\,VI]}}}
       {A_{\mathrm{Si}}\;f_{\mathrm{[Si\,VI]}}}
\end{equation}

The total coronal gas mass is given as:
\begin{equation}
M_{\mathrm{cor}}
= m_{p}\int n_{\mathrm{H}}\,dV
= m_{p}\int
  \frac{n_{\mathrm{[Si\,VI]}}}
       {A_{\mathrm{Si}}\;f_{\mathrm{[Si\,VI]}}}
\,dV
\end{equation}

Multiplying the numerator and denominator with constants $ n_{e}\epsilon_{\mathrm{[Si\,VI]}}$:
\begin{equation}
M_{\mathrm{cor}}
= m_{p}
  \int
    \frac{n_{e}\,n_{\mathrm{[Si\,VI]}}\,\epsilon_{\mathrm{[Si\,VI]}}}
         {A_{\mathrm{Si}}\;f_{\mathrm{[Si\,VI]}}\;n_{e}\;\epsilon_{\mathrm{[Si\,VI]}}}
  \,dV
= \frac{m_{p}}
         {A_{\mathrm{Si}}\;f_{\mathrm{[Si\,VI]}}\;n_{e}\;\epsilon_{\mathrm{[Si\,VI]}}}
  \int n_{e}\,n_{\mathrm{[Si\,VI]}}\,\epsilon_{\mathrm{[Si\,VI]}}\,dV
\end{equation}

Replacing the term under the integral with A1 and A2.

\begin{equation}
\boxed{
M_{\mathrm{cor}}
= \frac{m_{p}\,L_{\mathrm{[Si\,VI]}}}
         {A_{\mathrm{Si}}\;f_{\mathrm{[Si\,VI]}}\;n_{e}\;\epsilon_{\mathrm{[Si\,VI]}}}
}
\end{equation}
where the values of constants are\\
\begin{itemize}
    \item Mass of the proton: $m_p = 1.67 \times 10^{-24}\,\mathrm{g} $
    \item The abundance ratio is obtained from the \citealt{2009ARA&A..47..481A} definition, given as log $\epsilon$(Si) = log$_{10}$(N$_{\mathrm{Si}}$/N$_{\mathrm{H}}$) + 12, where (N$_{\mathrm{Si}}$/N$_{\mathrm{H}}$) is the abundance ratio $A_{\mathrm{Si}}$. For solar abundances, $A_{\mathrm{Si,\,Z_\odot}}  = 3.24 \times 10^{-5}$. If we assume a scaled-solar composition, where Si/H scales linearly with metallicity Z/Z$_\odot$, then the abundance ratio will also scale linearly, $A_{\mathrm{Si,\,Z}} = (Z/Z_\odot) \times A_{\mathrm{Si,\,Z_\odot}}$. 
    \item The ionization fraction of [Si\,VI], $f_{\mathrm{[Si\,VI]}}$ is assumed to be 0.5. This is taken as a conservative upper limit and we find that for lower $f_{\mathrm{[Si\,VI]}}$, the change in the energetics is comparable to the errors listed in Table \ref{table:energetics}.
    \item $\epsilon_{\mathrm{[Si\,VI]}}$ is determined using the line emission code PyNeb \citep{2015A&A...573A..42L} using the CHIANTI atomic database \citep{1997A&AS..125..149D, 2021ApJ...909...38D} using an electron temperature of 20000 K (which is the average temperatures of the gas traced by CL \citep{2021ApJ...920...62N}) and electron density of 100 cm$^{-3}$:
    $ \epsilon_{\mathrm{[Si\,VI]}} = 3.26 \times 10^{-21}\,\mathrm{erg\,cm^{3}\,s^{-1}} $
\end{itemize}

The calculations of the mass, momentum, and kinetic energy outflow rates depend on the above derived ionized gas mass and the spatial extent of the outflows and are given in Section \ref{sec:Hot ionized gas outflows}.
\FloatBarrier

\section{Coronal line fluxes and FWHM} \label{appendix:fluxes}
The coronal line fluxes and the FWHM for all the four targets. We find the total line fluxes by summing over the flux values in the fitted spatial maps and also present the extinction-corrected values. For the lines which could not be detected with sufficient S/N ratio in the individual spaxels, we fit the emission line by integrating a spectrum over a 0.3\arcsec aperture radius and report the total integrated flux. The average FWHM is the average of the FWHM values in the fitted spatial maps, or the value of FWHM in the integrated spectrum. We determined that the significant contribution to the error in the measured fluxes were primarily from the stellar component fitted using various stellar templates, and we report the error based off the flux determined from the fit without the stellar template. The upper limits for non-detected lines were estimated from the local continuum noise. For each line, we assumed an unresolved Gaussian profile with a FWHM set equal to the average FWHM of the detected emission lines in the same spectrum. The continuum level was measured from nearby line-free regions, and the noise was quantified using the median absolute deviation of the residuals after continuum subtraction. This 1$\sigma$ noise level was then scaled to 3$\sigma$, and the resulting 3$\sigma$ integrated flux upper limits are reported in the final column of the tables.
\FloatBarrier

\begin{deluxetable*}{cccccc}
\tablecaption{Coronal line fluxes for J0842}
\tablehead{
\colhead{Coronal line (Wavelength)} &
\colhead{Flux} &
\colhead{Flux (Extinction corrected)} &
\colhead{FWHM} &
\colhead{Flux Upper-limits} &
\colhead{Notes} \\
\colhead{} &
\colhead{(10$^{-20}$ W m$^{-2}$)} &
\colhead{(10$^{-20}$ W m$^{-2}$)} &
\colhead{(km s$^{-1}$)} &
\colhead{(10$^{-20}$ W m$^{-2}$)} &
\colhead{}\\
\colhead{(1)} & \colhead{(2)} & \colhead{(3)} & \colhead{(4)} & \colhead{(5)} & \colhead{(6)}
}
\startdata
$[$S\,VIII$]$ (0.9911 $\mu$m) & ... & ... & ... & 0.19 & Non-detection \\
$[$S\,IX$]$ (1.2520 $\mu$m) & ... & ... & ... & 0.08 & Non-detection \\
$[$Si\,X$]$ (1.4301 $\mu$m) & ... & ... & ... & 1.07 & Non-detection \\
$[$S\,XI$]$ (1.9220 $\mu$m) & ... & ... & ... & 0.25 & Non-detection \\
$[$Si\,VI$]$ (1.9630 $\mu$m) & ... & ... & ... & 0.34 & Non-detection \\
$[$Ca\,VIII$]$ (2.3205 $\mu$m)& ... & ... & ... & 0.70 & Non-detection \\
$[$Si\,VII$]$ (2.4826 $\mu$m) & ... & ... & ... & 0.33 & Non-detection \\
$[$Si\,IX$]$ (2.5839 $\mu$m) & ... & ... & ... & 0.23 & Non-detection \\
$[$Mg\,VIII$]$ (3.0276 $\mu$m)& ... & ... & ... & 0.26 & Non-detection \\
$[$Ca\,IV$]$ (3.2061 $\mu$m) & ... & ... & ... & 0.14 & Non-detection \\
$[$Al\,VI$]$ (3.6593 $\mu$m) & ... & ... & ... & 0.23 & Non-detection \\
$[$Si\,IX$]$ (3.9357 $\mu$m) & ... & ... & ... & 1.19 & Non-detection \\
$[$Ca\,VII$]$ (4.0864 $\mu$m) & ... & ... & ... & 0.18 & Non-detection \\
$[$Ca\,V$]$ (4.1574 $\mu$m) & ... & ... & ... & 0.25 & Non-detection \\
$[$Mg\,IV$]$ (4.4871 $\mu$m) & 1.01 $\pm$ 0.05 & 1.01 & 264 $\pm$ 66 & ... & ... \\
$[$Ar\,VI$]$ (4.5280 $\mu$m) & ... & ... & ... & 0.26 & Non-detection \\
$[$Na\,VII$]$ (4.6834 $\mu$m) & ... & ... & ... & 0.43 & Non-detection \\
\enddata
\tablecomments{\footnotesize
Column (1): Coronal line species and rest wavelength.
Column (2): Observed flux.
Column (3): Extinction-corrected flux.
Column (4): Average FWHM.
Column (5): Upper limits for non-detections.
Column (6): Notes on detections.
}
\end{deluxetable*}\label{tab:J0842}

\begin{deluxetable*}{cccccc}
\label{tab:J0906}
\tablecaption{Coronal line fluxes for J0906}
\tablehead{
\colhead{Coronal line (Wavelength)} &
\colhead{Flux} &
\colhead{Flux (Extinction corrected)} &
\colhead{FWHM} &
\colhead{Flux Upper-limits} &
\colhead{Notes} \\
\colhead{} &
\colhead{(10$^{-20}$ W m$^{-2}$)} &
\colhead{(10$^{-20}$ W m$^{-2}$)} &
\colhead{(km s$^{-1}$)} &
\colhead{(10$^{-20}$ W m$^{-2}$)} &
\colhead{}\\
\colhead{(1)} & \colhead{(2)} & \colhead{(3)} & \colhead{(4)} & \colhead{(5)} & \colhead{(6)}
}
\startdata
$[$S\,VIII$]$ (0.9911 $\mu$m) & 9.89 $\pm$ 2.34 & 11.8 & 334 $\pm$ 34 & ... & ... \\
$[$S\,IX$]$ (1.2520 $\mu$m) & 8.39 $\pm$ 2.50 & 10.0 & 315 $\pm$ 37 & ... & ... \\
$[$Si\,X$]$ (1.4301 $\mu$m) & 8.12 $\pm$ 0.99 & 9.71 & 281 $\pm$ 30 & ... & ... \\
$[$S\,XI$]$ (1.9220 $\mu$m) & 6.80 $\pm$ 1.54 & 8.13 & 455 $\pm$ 11 & ... & Int. spectrum \\
$[$Si\,VI$]$ (1.9630 $\mu$m) & 21.90 $\pm$ 6.52& 26.5 & 311 $\pm$ 40 & ... & ... \\
$[$Ca\,VIII$]$ (2.3205 $\mu$m) & ... & ... & ... & ... & Detector gap \\
$[$Si\,VII$]$ (2.4826 $\mu$m) & 19.68 $\pm$ 8.67& 23.5 & 320 $\pm$ 52 & ... & ... \\
$[$Si\,IX$]$ (2.5839 $\mu$m) & ... & ... & ... & 0.65 & Non-detection \\
$[$Mg\,VIII$]$ (3.0276 $\mu$m) & 27.14 $\pm$ 6.63& 32.5 & 338 $\pm$ 46 & ... & ... \\
$[$Ca\,IV$]$ (3.2061 $\mu$m) & 15.30 $\pm$ 0.47& 18.3 & 398 $\pm$ 20 & ... & Int. spectrum \\
$[$Al\,VI$]$ (3.6593 $\mu$m) & ... & ... & ... & 2.66 & Non-detection \\
$[$Si\,IX$]$ (3.9357 $\mu$m) & ... & ... & ... & ... & Detector gap \\
$[$Ca\,VII$]$ (4.0864 $\mu$m) & ... & ... & ... & 2.61 & Non-detection \\
$[$Ca\,V$]$ (4.1574 $\mu$m) & 5.15 $\pm$ 3.52 & 6.16 & 140 $\pm$ 17 & ... & Int. spectrum \\
$[$Mg\,IV$]$ (4.4871 $\mu$m) & 26.40 $\pm$ 4.77& 31.6 & 402 $\pm$ 15 & ... & Int. spectrum \\
$[$Ar\,VI$]$ (4.5280 $\mu$m) & 25.00 $\pm$ 5.78& 29.9 & 431 $\pm$ 18 & ... & Int. spectrum \\
$[$Na\,VII$]$ (4.6834 $\mu$m) & ... & ... & ... & 1.52 & Non-detection \\
\enddata
\tablecomments{\footnotesize
Same as Table \ref{tab:J0842}, but for J0906. The ‘Int. spectrum’ note indicates that the line is detected only in the integrated spectrum extracted from the central 0.3$\arcsec$ aperture.
}
\end{deluxetable*}

\begin{deluxetable*}{cccccc}
\tablecaption{Coronal emission line fluxes for J0954}
\tablehead{
\colhead{Coronal line (Wavelength)} &
\colhead{Flux} &
\colhead{Flux (Extinction corrected)} &
\colhead{FWHM} &
\colhead{Flux Upper-limits} &
\colhead{Notes} \\
\colhead{} &
\colhead{(10$^{-20}$ W m$^{-2}$)} &
\colhead{(10$^{-20}$ W m$^{-2}$)} &
\colhead{(km s$^{-1}$)} &
\colhead{(10$^{-20}$ W m$^{-2}$)} &
\colhead{}\\
\colhead{(1)} & \colhead{(2)} & \colhead{(3)} & \colhead{(4)} & \colhead{(5)} & \colhead{(6)}
}
\startdata
$[$S\,VIII$]$ (0.9911 $\mu$m) & 6.45 $\pm$ 0.46 & 7.61 & 215 $\pm$ 57 & ... & ... \\
$[$S\,IX$]$ (1.2520 $\mu$m) & 3.32 $\pm$ 2.09 & 3.92 & 155 $\pm$ 37 & ... & ... \\
$[$Si\,X$]$ (1.4301 $\mu$m) & 1.12 $\pm$ 1.00 & 1.32 & 63 $\pm$ 89 & ... & ... \\
$[$S\,XI$]$ (1.9220 $\mu$m) & ... & ... & ... & 0.08 & Non-detection \\
$[$Si\,VI$]$ (1.9630 $\mu$m) & 14.71 $\pm$ 2.88& 17.40& ... & ... & ... \\
\hspace{5mm} (Narrow) & 10.30 $\pm$ 2.58& 12.20& 127 $\pm$ 36 & ... & ... \\
\hspace{5mm} (Broad) & 4.41 $\pm$ 0.30& 5.20 & 374 $\pm$ 40 & ... & ... \\
$[$Ca\,VIII$]$ (2.3205 $\mu$m) & 3.16 $\pm$ 1.01 & 3.73 & 167 $\pm$ 13 & ... & Int. spectrum \\
$[$Si\,VII$]$ (2.4826 $\mu$m) & 22.10 $\pm$ 7.54& 26.00& ... & ... & ... \\
\hspace{5mm} (Narrow) & 12.00 $\pm$ 3.77& 14.10& 114 $\pm$ 37 & ... & ... \\
\hspace{5mm} (Broad) & 10.10 $\pm$ 3.77& 11.90& 396 $\pm$ 51 & ... & ... \\
$[$Si\,IX$]$ (2.5839 $\mu$m) & 4.85 $\pm$ 3.37 & 5.73 & 221 $\pm$ 45 & ... & ... \\
$[$Mg\,VIII$]$ (3.0276 $\mu$m) & 16.65 $\pm$ 5.76& 19.68& ... & ... & ... \\
\hspace{5mm} (Narrow) & 10.90 $\pm$ 4.83& 12.90& 185 $\pm$ 45 & ... & ... \\
\hspace{5mm} (Broad) & 5.75 $\pm$ 2.73& 6.78 & 375 $\pm$ 47 & ... & ... \\
$[$Ca\,IV$]$ (3.2061 $\mu$m) & 7.20 $\pm$ 2.17 & 8.50 & 177 $\pm$ 52 & ... & ... \\
$[$Al\,VI$]$ (3.6593 $\mu$m) & 2.44 $\pm$ 2.15 & 2.88 & 226 $\pm$ 68 & ... & ... \\
$[$Si\,IX$]$ (3.9357 $\mu$m) & ... & ... & ... & ... & Detector gap \\
$[$Ca\,VII$]$ (4.0864 $\mu$m) & 1.21 $\pm$ 0.38 & 1.43 & 173 $\pm$ 71 & ... & ... \\
$[$Ca\,V$]$ (4.1574 $\mu$m) & 5.10 $\pm$ 2.04 & 6.02 & 158 $\pm$ 48 & ... & ... \\
$[$Mg\,IV$]$ (4.4871 $\mu$m) & 12.81 $\pm$ 3.87& 15.05& ... & ... & ... \\
\hspace{5mm} (Narrow) & 10.40 $\pm$ 3.67& 12.20& 114 $\pm$ 39 & ... & ... \\
\hspace{5mm} (Broad) & 2.41 $\pm$ 0.20& 2.85 & 242 $\pm$ 63 & ... & ... \\
$[$Ar\,VI$]$ (4.5280 $\mu$m) & 27.02 $\pm$ 5.38& 31.96& 60 $\pm$ 26 & ... & ... \\
\hspace{5mm} (Narrow) & 21.80 $\pm$ 2.69& 25.80& 60 $\pm$ 26 & ... & ... \\
\hspace{5mm} (Broad) & 5.22 $\pm$ 2.69& 6.16 & 198 $\pm$ 48 & ... & ... \\
$[$Na\,VII$]$ (4.6834 $\mu$m) & ... & ... & ... & 1.70 & Non-detection \\
\enddata
\tablecomments{\footnotesize
Column (1): Coronal line species and rest wavelength.
Column (2): Observed flux.
Column (3): Extinction-corrected flux.
Column (4): Average FWHM.
Column (5): Upper limits for non-detections.
Column (6): Notes on special cases (non-detections, detector gaps, or integrated spectra. The ‘Int. spectrum’ note indicates that the line is detected only in the integrated spectrum extracted from the central 0.3$\arcsec$ aperture).
For multi-component fits, Narrow and Broad entries are listed separately beneath the total flux row.
}
\end{deluxetable*}\label{tab:J0954_broad}

\begin{deluxetable*}{cccccc}
\tablecaption{Coronal emission line fluxes for J1009}
\tablehead{
\colhead{Coronal line (Wavelength)} & 
\colhead{Flux} & 
\colhead{Flux (Extinction corrected)} & 
\colhead{FWHM} & 
\colhead{Flux Upper-limits} & 
\colhead{Notes} \\
\colhead{} & 
\colhead{(10$^{-20}$ W m$^{-2}$)} & 
\colhead{(10$^{-20}$ W m$^{-2}$)} & 
\colhead{(km s$^{-1}$)} & 
\colhead{(10$^{-20}$ W m$^{-2}$)} & 
\colhead{}\\
\colhead{(1)} & \colhead{(2)} & \colhead{(3)} & \colhead{(4)} & \colhead{(5)} & \colhead{(6)}
}
\startdata
$[$S\,VIII$]$ (0.9911 $\mu$m)   & 6.36 $\pm$ 0.81 & 6.85 & 243 $\pm$ 58 & ...   & ... \\
$[$S\,IX$]$ (1.2520 $\mu$m)     & 4.82 $\pm$ 0.56 & 5.18 & 124 $\pm$ 38 & ...   & ... \\
$[$Si\,X$]$ (1.4301 $\mu$m)     & ...             & ...  & ...          & ...   & Detector gap \\
$[$S\,XI$]$ (1.9220 $\mu$m)     & 1.19 $\pm$ 0.58 & 1.28 & 192 $\pm$ 49 & ...   & ... \\
$[$Si\,VI$]$ (1.9630 $\mu$m)    & 14.45 $\pm$ 2.60& 15.48& ...          & ...   & ... \\
\hspace{5mm} (Narrow)            & 12.80 $\pm$ 2.30& 13.70& 134 $\pm$ 37 & ...   & ... \\
\hspace{5mm} (Broad)             & 1.65  $\pm$ 0.30& 1.78 & 317 $\pm$ 32 & ...   & ... \\
$[$Ca\,VIII$]$ (2.3205 $\mu$m)  & 1.87 $\pm$ 0.29 & 2.01 & 150 $\pm$ 38 & ...   & ... \\
$[$Si\,VII$]$ (2.4826 $\mu$m)   & 19.78 $\pm$ 3.63& 21.26& ...          & ...   & ... \\
\hspace{5mm} (Narrow)            & 17.40 $\pm$ 2.28& 18.70& 102 $\pm$ 28 & ...   & ... \\
\hspace{5mm} (Broad)             & 2.38  $\pm$ 1.35& 2.56 & 466 $\pm$ 106& ...   & ... \\
$[$Si\,IX$]$ (2.5839 $\mu$m)    & 3.90 $\pm$ 0.83 & 4.20 & 133 $\pm$ 31 & ...   & ... \\
$[$Mg\,VIII$]$ (3.0276 $\mu$m)  & 17.06 $\pm$ 1.54& 18.33& ...          & ...   & ... \\
\hspace{5mm} (Narrow)            & 10.80 $\pm$ 1.53& 11.60& 170 $\pm$ 48 & ...   & ... \\
\hspace{5mm} (Broad)             & 6.26  $\pm$ 0.10& 6.73 & 378 $\pm$ 90 & ...   & ... \\
$[$Ca\,IV$]$ (3.2061 $\mu$m)    & 1.54 $\pm$ 0.55 & 1.66 & 142 $\pm$ 28 & ...   & ... \\
$[$Al\,VI$]$ (3.6593 $\mu$m)    & 1.90 $\pm$ 0.31 & 2.05 & 281 $\pm$ 74 & ...   & ... \\
$[$Si\,IX$]$ (3.9357 $\mu$m)    & 9.77 $\pm$ 0.50 & 11.52& ...          & ...   & ... \\
\hspace{5mm} (Narrow)            & 8.82  $\pm$ 0.25& 9.49 & 102 $\pm$ 28 & ...   & ... \\
\hspace{5mm} (Broad)             & 0.95  $\pm$ 0.25& 1.03 & 466 $\pm$ 106& ...   & ... \\
$[$Ca\,VII$]$ (4.0864 $\mu$m)   & 0.58 $\pm$ 0.27 & 0.70 & 186 $\pm$ 129& ...   & ... \\
$[$Ca\,V$]$ (4.1574 $\mu$m)     & 2.20 $\pm$ 0.26 & 2.37 & 143 $\pm$ 28 & ...   & ... \\
$[$Mg\,IV$]$ (4.4871 $\mu$m)    & 12.70 $\pm$ 6.40& 13.70& 97  $\pm$ 22 & ...   & ... \\
$[$Ar\,VI$]$ (4.5280 $\mu$m)    & 45.90 $\pm$ 11.0& 49.46& ...          & ...   & ... \\
\hspace{5mm} (Narrow)            & 43.80 $\pm$ 10.4& 47.20& 106 $\pm$ 23 & ...   & ... \\
\hspace{5mm} (Broad)             & 2.10  $\pm$ 0.60& 2.26 & 414 $\pm$ 72 & ...   & ... \\
$[$Na\,VII$]$ (4.6834 $\mu$m)   & 0.83 $\pm$ 0.28 & 0.89 & 136 $\pm$ 34 & ...   & ... \\
\enddata
\tablecomments{\footnotesize 
Same as Table \ref{tab:J0954_broad}, but for J1009.
}
\label{tab:J1009}
\end{deluxetable*}

\FloatBarrier
\bibliography{JWST_CL}{}
\bibliographystyle{aasjournalv7}



\end{document}